\pdfoutput=1
\documentclass{JINST}
\usepackage{graphicx,amssymb,subfig}
\usepackage{amsmath}
\usepackage{caption}
\usepackage{url}
\usepackage{cite}
\usepackage{lineno}
\usepackage{microtype}
\usepackage{acronym}

\DeclareMathAlphabet{\mathcal}{OMS}{cmsy}{m}{n}
\newcommand{\figwidth}{0.65}
\newcommand{\figwidthS}{0.5}

\title{Antennas for the Detection of Radio Emission Pulses from Cosmic-Ray induced Air Showers at the Pierre Auger Observatory}

\author{ 
\par\noindent
P.~Abreu$^{63}$, 
M.~Aglietta$^{51}$, 
M.~Ahlers$^{94}$, 
E.J.~Ahn$^{81}$, 
I.F.M.~Albuquerque$^{15}$, 
D.~Allard$^{29}$, 
I.~Allekotte$^{1}$, 
J.~Allen$^{85}$, 
P.~Allison$^{87}$, 
A.~Almela$^{11,\: 7}$
% ...
\begin{center}
(The Pierre Auger Collaboration)      
\vskip 0.75cm
and
D.~Charrier$^{33}$,
L.~Denis$^{j}$,
G.~Hilgers$^{39}$,
L.~Mohrmann$^{39}$,
B.~Philipps$^{39}$,
O.~Seeger$^{39}$
\end{center}

\vskip 0.5cm
\begin{flushleft}
\begin{itshape}
\begin{small}
\par\noindent
$^{1}$ Centro At\'{o}mico Bariloche and Instituto Balseiro (CNEA-UNCuyo-CONICET), San 
Carlos de Bariloche, 
Argentina \\
$^{2}$ Centro de Investigaciones en L\'{a}seres y Aplicaciones, CITEDEF and CONICET, 
Argentina \\
$^{3}$ Departamento de F\'{\i}sica, FCEyN, Universidad de Buenos Aires y CONICET, 
Argentina \\
$^{4}$ IFLP, Universidad Nacional de La Plata and CONICET, La Plata, 
Argentina \\
$^{5}$ Instituto de Astronom\'{\i}a y F\'{\i}sica del Espacio (CONICET-UBA), Buenos Aires, 
Argentina \\
$^{6}$ Instituto de F\'{\i}sica de Rosario (IFIR) - CONICET/U.N.R. and Facultad de Ciencias 
Bioqu\'{\i}micas y Farmac\'{e}uticas U.N.R., Rosario, 
Argentina \\
$^{7}$ Instituto de Tecnolog\'{\i}as en Detecci\'{o}n y Astropart\'{\i}culas (CNEA, CONICET, UNSAM), 
Buenos Aires, 
Argentina \\
$^{8}$ National Technological University, Faculty Mendoza (CONICET/CNEA), Mendoza, 
Argentina \\
$^{9}$ Observatorio Pierre Auger, Malarg\"{u}e, 
Argentina \\
$^{10}$ Observatorio Pierre Auger and Comisi\'{o}n Nacional de Energ\'{\i}a At\'{o}mica, Malarg\"{u}e, 
Argentina \\
$^{11}$ Universidad Tecnol\'{o}gica Nacional - Facultad Regional Buenos Aires, Buenos Aires,
Argentina \\
$^{12}$ University of Adelaide, Adelaide, S.A., 
Australia \\
$^{13}$ Centro Brasileiro de Pesquisas Fisicas, Rio de Janeiro, RJ, 
Brazil \\
$^{14}$ Universidade de S\~{a}o Paulo, Instituto de F\'{\i}sica, S\~{a}o Carlos, SP, 
Brazil \\
$^{15}$ Universidade de S\~{a}o Paulo, Instituto de F\'{\i}sica, S\~{a}o Paulo, SP, 
Brazil \\
$^{16}$ Universidade Estadual de Campinas, IFGW, Campinas, SP, 
Brazil \\
$^{17}$ Universidade Estadual de Feira de Santana, 
Brazil \\
$^{18}$ Universidade Estadual do Sudoeste da Bahia, Vitoria da Conquista, BA, 
Brazil \\
$^{19}$ Universidade Federal da Bahia, Salvador, BA, 
Brazil \\
$^{20}$ Universidade Federal do ABC, Santo Andr\'{e}, SP, 
Brazil \\
$^{21}$ Universidade Federal do Rio de Janeiro, Instituto de F\'{\i}sica, Rio de Janeiro, RJ, 
Brazil \\
$^{22}$ Universidade Federal Fluminense, EEIMVR, Volta Redonda, RJ, 
Brazil \\
$^{23}$ Rudjer Bo\v{s}kovi\'c Institute, 10000 Zagreb, 
Croatia \\
$^{24}$ Charles University, Faculty of Mathematics and Physics, Institute of Particle and 
Nuclear Physics, Prague, 
Czech Republic \\
$^{25}$ Institute of Physics of the Academy of Sciences of the Czech Republic, Prague, 
Czech Republic \\
\end{small}
\end{itshape}
\end{flushleft}
}
\author{ 
\begin{flushleft}
\begin{itshape}
\begin{small}
\par\noindent
$^{26}$ Palacky University, RCPTM, Olomouc, 
Czech Republic \\
$^{28}$ Institut de Physique Nucl\'{e}aire d'Orsay (IPNO), Universit\'{e} Paris 11, CNRS-IN2P3, 
Orsay, 
France \\
$^{29}$ Laboratoire AstroParticule et Cosmologie (APC), Universit\'{e} Paris 7, CNRS-IN2P3, 
Paris, 
France \\
$^{30}$ Laboratoire de l'Acc\'{e}l\'{e}rateur Lin\'{e}aire (LAL), Universit\'{e} Paris 11, CNRS-IN2P3, 
France \\
$^{31}$ Laboratoire de Physique Nucl\'{e}aire et de Hautes Energies (LPNHE), Universit\'{e}s 
Paris 6 et Paris 7, CNRS-IN2P3, Paris, 
France \\
$^{32}$ Laboratoire de Physique Subatomique et de Cosmologie (LPSC), Universit\'{e} Joseph
 Fourier, INPG, CNRS-IN2P3, Grenoble, 
France \\
$^{33}$ SUBATECH, \'{E}cole des Mines de Nantes, CNRS-IN2P3, Universit\'{e} de Nantes, 
France \\
$^{34}$ Bergische Universit\"{a}t Wuppertal, Wuppertal, 
Germany \\
$^{35}$ Karlsruhe Institute of Technology - Campus North - Institut f\"{u}r Kernphysik, Karlsruhe, 
Germany \\
$^{36}$ Karlsruhe Institute of Technology - Campus North - Institut f\"{u}r 
Prozessdatenverarbeitung und Elektronik, Karlsruhe, 
Germany \\
$^{37}$ Karlsruhe Institute of Technology - Campus South - Institut f\"{u}r Experimentelle 
Kernphysik (IEKP), Karlsruhe, 
Germany \\
$^{38}$ Max-Planck-Institut f\"{u}r Radioastronomie, Bonn, 
Germany \\
$^{39}$ RWTH Aachen University, III. Physikalisches Institut A, Aachen, 
Germany \\
$^{40}$ Universit\"{a}t Hamburg, Hamburg, 
Germany \\
$^{41}$ Universit\"{a}t Siegen, Siegen, 
Germany \\
$^{42}$ Dipartimento di Fisica dell'Universit\`{a} and INFN, Genova, 
Italy \\
$^{43}$ Universit\`{a} dell'Aquila and INFN, L'Aquila, 
Italy \\
$^{44}$ Universit\`{a} di Milano and Sezione INFN, Milan, 
Italy \\
$^{45}$ Universit\`{a} di Napoli "Federico II" and Sezione INFN, Napoli, 
Italy \\
$^{46}$ Universit\`{a} di Roma II "Tor Vergata" and Sezione INFN,  Roma, 
Italy \\
$^{47}$ Universit\`{a} di Catania and Sezione INFN, Catania, 
Italy \\
$^{48}$ Universit\`{a} di Torino and Sezione INFN, Torino, 
Italy \\
$^{49}$ Dipartimento di Matematica e Fisica "E. De Giorgi" dell'Universit\`{a} del Salento and 
Sezione INFN, Lecce, 
Italy \\
$^{50}$ Istituto di Astrofisica Spaziale e Fisica Cosmica di Palermo (INAF), Palermo, 
Italy \\
$^{51}$ Istituto di Fisica dello Spazio Interplanetario (INAF), Universit\`{a} di Torino and 
Sezione INFN, Torino, 
Italy \\
$^{52}$ INFN, Laboratori Nazionali del Gran Sasso, Assergi (L'Aquila), 
Italy \\
$^{53}$ Benem\'{e}rita Universidad Aut\'{o}noma de Puebla, Puebla, 
Mexico \\
$^{54}$ Centro de Investigaci\'{o}n y de Estudios Avanzados del IPN (CINVESTAV), M\'{e}xico, 
Mexico \\
$^{55}$ Universidad Michoacana de San Nicolas de Hidalgo, Morelia, Michoacan, 
Mexico \\
$^{56}$ Universidad Nacional Autonoma de Mexico, Mexico, D.F., 
Mexico \\
$^{57}$ IMAPP, Radboud University Nijmegen, 
Netherlands \\
$^{58}$ Kernfysisch Versneller Instituut, University of Groningen, Groningen, 
Netherlands \\
$^{59}$ Nikhef, Science Park, Amsterdam, 
Netherlands \\
$^{60}$ ASTRON, Dwingeloo, 
Netherlands \\
$^{61}$ Institute of Nuclear Physics PAN, Krakow, 
Poland \\
$^{62}$ University of \L \'{o}d\'{z}, \L \'{o}d\'{z}, 
Poland \\
$^{63}$ LIP and Instituto Superior T\'{e}cnico, Technical University of Lisbon, 
Portugal \\
$^{64}$ 'Horia Hulubei' National Institute for Physics and Nuclear Engineering, Bucharest-
Magurele, 
Romania \\
\end{small}
\end{itshape}
\end{flushleft}
}
\author{ 
\begin{flushleft}
\begin{itshape}
\begin{small}
\par\noindent
$^{65}$ University of Bucharest, Physics Department, 
Romania \\
$^{66}$ University Politehnica of Bucharest, 
Romania \\
$^{67}$ J. Stefan Institute, Ljubljana, 
Slovenia \\
$^{68}$ Laboratory for Astroparticle Physics, University of Nova Gorica, 
Slovenia \\
$^{69}$ Instituto de F\'{\i}sica Corpuscular, CSIC-Universitat de Val\`{e}ncia, Valencia, 
Spain \\
$^{70}$ Universidad Complutense de Madrid, Madrid, 
Spain \\
$^{71}$ Universidad de Alcal\'{a}, Alcal\'{a} de Henares (Madrid), 
Spain \\
$^{72}$ Universidad de Granada \&  C.A.F.P.E., Granada, 
Spain \\
$^{73}$ Universidad de Santiago de Compostela, 
Spain \\
$^{74}$ Rudolf Peierls Centre for Theoretical Physics, University of Oxford, Oxford, 
United Kingdom \\
$^{75}$ School of Physics and Astronomy, University of Leeds, 
United Kingdom \\
$^{76}$ Argonne National Laboratory, Argonne, IL, 
USA \\
$^{77}$ Case Western Reserve University, Cleveland, OH, 
USA \\
$^{78}$ Colorado School of Mines, Golden, CO, 
USA \\
$^{79}$ Colorado State University, Fort Collins, CO, 
USA \\
$^{80}$ Colorado State University, Pueblo, CO, 
USA \\
$^{81}$ Fermilab, Batavia, IL, 
USA \\
$^{82}$ Los Alamos National Laboratory, Los Alamos, NM, 
USA \\
$^{83}$ Louisiana State University, Baton Rouge, LA, 
USA \\
$^{84}$ Michigan Technological University, Houghton, MI, 
USA \\
$^{85}$ New York University, New York, NY, 
USA \\
$^{86}$ Northeastern University, Boston, MA, 
USA \\
$^{87}$ Ohio State University, Columbus, OH, 
USA \\
$^{88}$ Pennsylvania State University, University Park, PA, 
USA \\
$^{89}$ Southern University, Baton Rouge, LA, 
USA \\
$^{90}$ University of Chicago, Enrico Fermi Institute, Chicago, IL, 
USA \\
$^{91}$ University of Hawaii, Honolulu, HI, 
USA \\
$^{92}$ University of Nebraska, Lincoln, NE, 
USA \\
$^{93}$ University of New Mexico, Albuquerque, NM, 
USA \\
$^{94}$ University of Wisconsin, Madison, WI, 
USA \\
$^{95}$ University of Wisconsin, Milwaukee, WI, 
USA \\
$^{96}$ Institute for Nuclear Science and Technology (INST), Hanoi, 
Vietnam \\
\par\noindent
(\ddag) Deceased \\
(a) at Konan University, Kobe, Japan \\
(b) now at the Universidad Autonoma de Chiapas on leave of absence from Cinvestav \\
(f) now at University of Maryland \\
(h) now at NYU Abu Dhabi \\
(i) now at Universit\'{e} de Lausanne \\
(j) Station de Radioastronomie de Nan\c{c}ay, Observatoire de Paris, Nan\c{c}ay, France
\end{small}
\end{itshape}
\end{flushleft}
}

\abstract{The Pierre Auger Observatory is exploring the potential of the radio detection technique to study extensive air showers induced by ultra-high energy cosmic rays.
The Auger Engineering Radio Array (AERA) addresses both technological and scientific aspects of the radio technique. A first phase of AERA has been operating since September 2010 with detector stations observing radio signals at frequencies between 30 and 80 MHz. In this paper we present comparative studies to identify and optimize the antenna design for the final configuration of AERA consisting of 160 individual radio detector stations. 
The transient nature of the air shower signal requires a detailed description of the antenna sensor.
As the ultra-wideband reception of pulses is not widely discussed in antenna literature,
we review the relevant antenna characteristics and enhance theoretical considerations towards the impulse response of antennas including polarization effects and multiple signal reflections.
On the basis of the vector effective length we study the transient response characteristics of three candidate antennas in the time domain. 
Observing the variation of the continuous galactic background intensity we rank the antennas with respect to the noise level added to the galactic signal.
}


\begin{document}

  \section{Introduction}

The Pierre Auger Observatory \cite{PropsPAO2004} is a hybrid detector for the observation of cosmic rays above $\sim 10^{18}$ eV. The atmosphere of the Earth is used as a calorimeter measuring the particle shower that evolves after the penetration of a primary cosmic ray. 
These air showers are observed with an array of 1660 ground based particle detectors covering an area of $~ 3000$ km$^2$. 
The array is overlooked by 27 optical telescopes \cite{Abraham2010227} which are sensitive to the fluorescence light emitted by nitrogen molecules which were excited by the charged particles of the passing air shower. 
The combination of both detection techniques allows for a precise determination of the energy and arrival direction of cosmic rays and gives information on the chemical composition of the cosmic ray flux \cite{Abraham2010, PhysRevLett.104.091101}.

Besides the established detection techniques, the Pierre Auger Collaboration is exploring the possibility of detecting extensive air showers via radio pulses that are generated as the showers develop in the atmosphere \cite{Kelley2011b,Revenu2011,Allison2011}.
The radio signal strength promises a quadratic scaling with the energy of the cosmic ray, high angular resolution in the reconstruction of the air shower axis, and sensitivity to the nature of the primary particle. In combination with an almost 100\% duty cycle these attributes make a radio system a candidate for the next generation of ground-based air shower detectors.

The emission of electromagnetic radiation from air showers in the MHz frequency regime was first observed by Jelley and co-workers in 1965 \cite{Jelley1965}. It was found that air showers emit an electromagnetic pulse in the direction of the shower propagation. The observation of the wavefront with an array of individual antennas at different positions with respect to the shower axis should allow a reconstruction of the properties of the air shower and the corresponding primary cosmic ray. In the following years, progress was made with experiments reporting air-shower observations in a frequency range from 2 to 550 MHz \cite{Allan1970a,Fegan1973}. The realization of a comprehensive radio detector, however, was not feasible until the appearance of fast digital oscilloscopes in the past two decades. Since 2005 the CODALEMA \cite{Ardouin2005} and LOPES \cite{Falcke2005} experiments have succeed in detecting air showers up to energies of $10^{18}$ eV. They confirmed the predicted importance of the Earth's 
magnetic field for the generation process of the radio pulse \cite{Falcke2003,Huege2003}.

\begin{figure}
  \centering
  \includegraphics[width=\figwidth\textwidth]{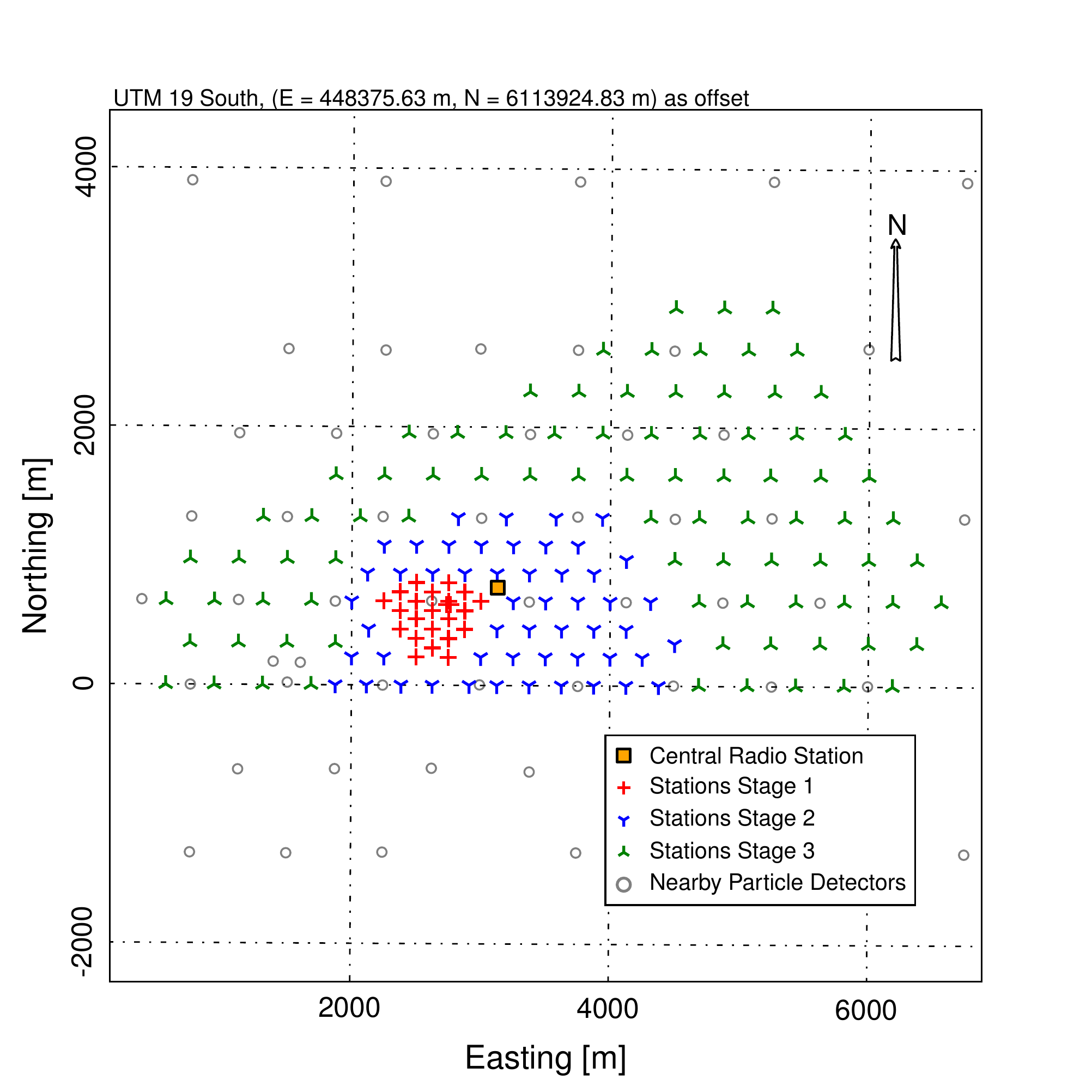}
  \caption{Layout of the Auger Engineering Radio Array with its radio detector stations embedded in the grid of particle detectors in the north-western part of the Pierre Auger Observatory. AERA has an extent of 
  $\sim20\,\mbox{km}^2 $
  and consists of 160 radio detector stations that will be deployed in three stages. The first stage of AERA is operating since September 2010. The Central Radio Station hosts the central data acquisition of AERA.}
  \label{fig:AERA}
\end{figure}

The Auger Engineering Radio Array (AERA) 
is a radio detector situated at the Pierre Auger Observatory. AERA will instrument a sensitive area of 20 km$^2$ with 160 detector stations and is thus the first detector 
with sufficient collecting area to make possible the
measurement of radio signals of air showers beyond $10^{18}$ eV.
The layout of AERA shown in Fig. \ref{fig:AERA}. It features a varying spacing between the detector stations which is intended to maximize the number of recorded events over a wide energy range from roughly $10^{17}$ to $10^{19}$ eV at a rate of several thousand air showers per year \cite{Fliescher2010}.
AERA is co-located with 
Auger fluorescence telescopes and its surface detector. 
Hence, the Pierre Auger Observatory offers the unique possibility to study radio emission 
even at large distances from the shower axis. 
The first stage of AERA consists of 21 autonomous detector stations forming the dense core seen in Fig. \ref{fig:AERA} and has been operating since September 2010.


AERA is a self-triggered radio array: at each detector station the radio signal is observed continuously and a trigger decision is formed to select air showers which produce a transient signal of a few times 10 ns in length. Individual trigger decisions are
collected in a data acquisition system where signal patterns are investigated that match air showers recorded in multiple detector stations.

\begin{figure}
  \centering
  \includegraphics[width=\figwidthS\textwidth]{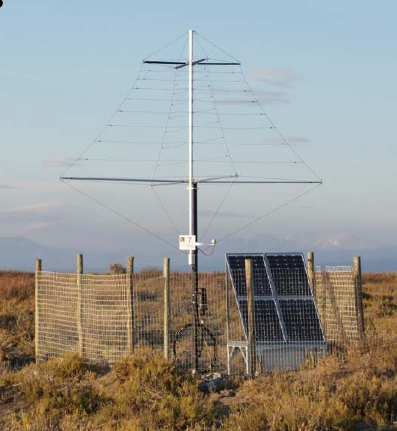}
  \caption{A radio detector station of the first stage of AERA. A logarithmic-periodic dipole antenna, called Small Black Spider, is used as sensor for the radio emission of extensive air showers. The antenna is read out with digital oscilloscopes and a trigger decision on air shower signals is performed by an FPGA. A photovoltaic power system enables autonomous operation of the detector station and supplies the readout electronics, the GPS system used for timing, and the communication system.}
  \label{fig:RDS}
\end{figure}

The radio detector stations provide ultra-wideband reception of radio signals from 30 to 80 MHz, a bandwidth locally free of AM and FM band transmitters. The continuous radio signal is sampled at a rate of 180 MHz and processed with a field-programmable gate array (FPGA) to form trigger decisions. A review of the readout electronics is given in Ref. \cite{Ruehle2010}. Fig. \ref{fig:RDS} shows a picture of a radio detector station.

The data recorded by the detector stations is a convolution of the radio signal and the response of the readout electronics. To create a measurement that is independent of the detector setup, the impact of the individual hardware components has to be removed from the AERA data. Here, the antenna deserves special attention as its frequency response is highly non-linear and depends on the incoming direction and polarization of the recorded signal. Hence, a precise knowledge and proper use of the relevant antenna characteristics is required to obtain a calibrated measurement of the radio emission from air showers.

The antenna and its amplifier
determine the signal to noise level that is obtained in air shower observations. An optimal noise performance of the antenna will maximize the sensitivity of the radio detector stations to air shower signals and the efficiency of the detector to cosmic rays.


In this article we present our studies to characterize and evaluate candidate antennas for the next stage of AERA. The article is structured as follows: in section 2 the candidate antennas are presented. In the third section we focus on the antenna theory needed for understanding the reception of transient signals. Having identified the relevant antenna characteristics we show in section 4 how to access them in simulation and measurement. In section 5 we compare the response of the tested antennas to transient signals on the basis of antenna simulations. 
Preferably, an antenna for the detection of air showers should introduce only minor distortions to the recorded signal shape. The successful observation of air showers will be determined by the signal to noise ratio obtained in measurements. Therefore we present in section 6 comparative measurements of the variation of the galactic noise level performed at the Nan\c{c}ay Radio Observatory, France, which allow us to discriminate the candidate antennas with respect to their noise performance.

  \section{Antennas for the Detection of Radio Signals from Cosmic-Ray Induced Air Showers}
\label{sec:Antennas}

In this section we present three antenna models that have been evaluated for the setup at AERA. The three investigated models address the task of radio detection of air showers with different concepts. All antennas presented are the result of several stages of development, taking into account the experience gained in smaller radio detection setups both in Europe and at the site of the Pierre Auger Observatory.
This is necessary as the environmental conditions of the Argentinian Pampa impose special demands on the antenna structure for instance through wind loads of up to 160 km/h. Hence, the durability and consequent costs for maintenance directly impact on the success of an antenna model, especially for a radio detector design with a large number of detector stations.

With respect to the electrical properties of the radio sensors it is useful to consider the antenna as an integral combination of the metallic structure capturing the signal and the first low noise amplifier (LNA). Whereas the structure determines the directional properties of the antenna, the ultra-wideband reception of the antenna is ruled by the combination of both elements.

Requirements on the directional properties of the antenna are imposed by the widespread layout of the radio detector array. At each detector station the full sky needs to be observed
so omni-directional antennas are used. 
To measure the polarization of the radio signal the placement of at least two perpendicular antennas is required at each detector station.

Currently, the properties of the radio pulse and its generation mechanism are subject to research beyond its capabilities as a tool to detect the cosmic ray. Hence the antennas have to be sensitive in a broad frequency range to allow for a maximum detail of the observation. At AERA the bandwidth is limited by the presence of AM band transmitters below 30 MHz and FM band transmitters mainly above 80 MHz.

\subsection{The Small Black Spider Antenna}
\label{sec:SBS}
Logarithmic periodic dipole antennas (LPDAs) are being used for the first stage of AERA. The logarithmic periodic principle assembles a series of half wave dipoles of increasing length to keep the radiation resistance of the antenna constant over a wide frequency range.
LPDAs were first adapted to the needs of radio detection for the LOPES-STAR experiment \cite{Gemmeke2006}.
The `Small Black Spider` realizes the LPDA principle as a wire antenna and is shown in Fig. \ref{fig:RDS}.
The 
antenna 
integrates two independent antenna planes in the same mechanical structure which has a dimension of $4\times4\times3.5$ m$^3$ and a weight of 18 kg. For transportation purposes the design of the Small Black Spider includes a folding mechanism in the antenna structure. This allows one to assemble the antenna completely in the laboratory and make it operational within 15 minutes at the detector site.

The lengths of the shortest and longest dipoles determine the available frequency range and have been adapted to the AERA frequency and for the Small Black Spider. A slightly enhanced sensitivity of the antenna to the top direction is obtained as the amplitudes fed from the individual dipoles into a common wave guide add up constructively at the footpoint where the antenna is read out. The footpoint of the Small Black Spider is at the top of the structure which is in a height of $\sim 4.5$ m when installed in the field. Although the footpoint is the optimum position for the amplifier such placement is not feasible due to maintenance constraints. Instead a matched feed of the footpoint into a $50\,\Omega$ coaxial cable is obtained using a transmission line transformer with a 4:1 impedance ratio. This is possible because of a constant antenna impedance of $200\,\Omega$ within the frequency range. The coaxial cable guides the signal to the LNA at the bottom of the antenna. Here, the LNA includes filter elements 
at its input to further ensure the frequency selectivity of the antenna to the AERA band. Details of the development of the LNA are presented in Ref. \cite{Stephan2010a}.

For the first stage of AERA, 30 Small Black Spiders have been produced including spares and antennas for test setups. Within one year of placement in the field the antennas have proven to be both robust against the environmental conditions of the Argentinian Pampa and sensitive to radio signals from cosmic ray induced air showers. A detailed description of the antenna is given in Ref. \cite{Seeger2010}.

\subsection{The Salla Antenna}
\begin{figure}
  \centering
  \includegraphics[width=\figwidthS\textwidth]{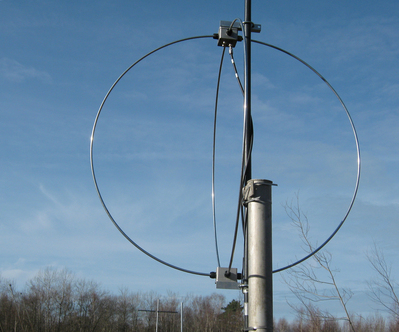}
  \caption{The Salla antenna during test measurements at the Nan\c{c}ay Radio Observatory. The Salla realizes a Beverage antenna as a dipole loop of 1.2 m diameter. The antenna is read out at the top with an LNA. The amplified signal is guided through the antenna structure to the bottom. }
  \label{fig:Salla}
\end{figure}
The short aperiodic loaded loop antenna 'Salla' realizes a Beverage antenna \cite{Beverage1941} as a dipole loop of 1.2 m diameter. 
The Salla has been developed to provide a minimal design that fulfills the need for both, ultra-wideband sensitivity, and low costs for production and maintenance of the antenna in a large scale radio detector. The compact structure of the Salla makes the antenna robust and easy to manufacture. A picture of a Salla showing its two polarization planes is displayed in Fig. \ref{fig:Salla}. 

Beverage antennas include a resistor load within the antenna structure to give a specific shape to the directivity. In the case of the Salla a resistance of $500\,\Omega$ connects the ends of the dipole arms at the bottom of the antenna. The antenna is read out at the top which is also the position of the LNA. While signals coming from above will induce a current directly at the input of the amplifier, the reception from directions below the antenna is strongly suppressed as the captured power is primarily consumed within the ohmic resistor rather than amplified by the LNA. The resulting strong suppression of sensitivity towards the ground reduces the dependence of the antenna on environmental conditions which might vary as a function of time and are thus a source of systematic uncertainty. With the inclusion of an ohmic resistor the Salla especially challenges its amplifier as only $\sim10\%$ of the captured signal intensity is available at the input of the LNA. Proper matching between the antenna structure 
and the LNA is realized with a 3:1 transmission line transformer. The structure of the Salla creates a sensitivity which is flat as a function of frequency.

Salla antennas are used at the Tunka radio detector \cite{Antokhonov2011}. They have been tested in a radio detection setup at the Pierre Auger Observatory and within the LOPES experiment. A detailed description of the Salla antenna is given in Ref. \cite{Gemmeke2009}.

\subsection{The Butterfly Antenna}
\label{sec:Butterfly}
\begin{figure}
  \centering
  \includegraphics[width=\figwidthS\textwidth]{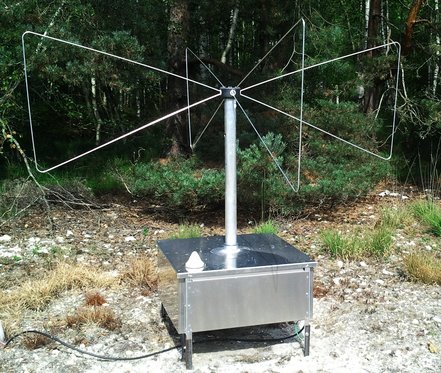}
  \caption{A Butterfly antenna installed at the site of the Nan\c{c}ay Radio Observatory. The antenna construction aims at a close integration with a box containing the readout electronics and a center pole which is also used to carry an antenna for wireless communication.}
  \label{fig:Butterfly}
\end{figure}

The 'Butterfly' is an active bowtie antenna and the successor of the active short dipole antenna \cite{Charrier2007} used for the CODALEMA radio detector. The Butterfly has dimensions of $2\times2\times1$ m$^3$ and is constructed with fat dipoles. The dipole signals are fed directly into the two input channels of an LNA at the center of the antenna which constitutes the active antenna concept. The dipoles are self-supporting and their hollow construction reduces the sensitivity to heavy wind loads. The Butterfly antenna is shown in Fig. \ref{fig:Butterfly}. 

The Butterfly explicitly uses the presence of the ground to enhance its signal. The center of the antenna is installed at a height of 1.5 m. Here, the direct wave and the wave reflected on the ground add constructively in the antenna throughout most combinations of frequency and incoming directions. Ultra-wideband sensitivity is obtained by designing the input impedance of the LNA to depend on the impedance of the fat dipole structure as a function of frequency. In this way, the sensitivity of the dipole has been optimized allowing for an efficient detection also of wavelengths that are much longer than the dimension of the antenna structure. The LNA itself is a application-specific integrated circuit (ASIC) and does not require a transformer since its input is differential \cite{Charrier2007}.

In the current extension of the CODALEMA experiment \cite{Ravel2011} 33 Butterfly antennas are deployed. The antenna was used successfully to observe cosmic rays in one of the pioneering setups at the Pierre Auger Observatory \cite{Revenu2010}. Details of the Butterfly antenna are given in Ref. \cite{Charrier2010}.


  \section{Antenna Theory}
\label{sec:AntennaTheory}
To perform a calibrated measurement of the radio emission from cosmic ray induced air showers the impact of the detector and especially of the antenna needs to be unfolded from the recorded signals. 

The goal of this section is to introduce theoretical aspects of ultra-wideband antennas needed to describe the interrelation between the measured voltage $V(t)$ responding to an incident electric field $\vec{E}(t)$. Here, we aim to unify the calculations for the diverse antennas described in Sec. \ref{sec:Antennas}. Having identified the relevant quantities, a closer evaluation of the antenna models will take place in the following sections of this article.

\subsection{The Vector Effective Length}
\label{sec:VectorEffectiveLength}
\begin{figure}
  \centering
  \includegraphics[width=\figwidthS\textwidth]{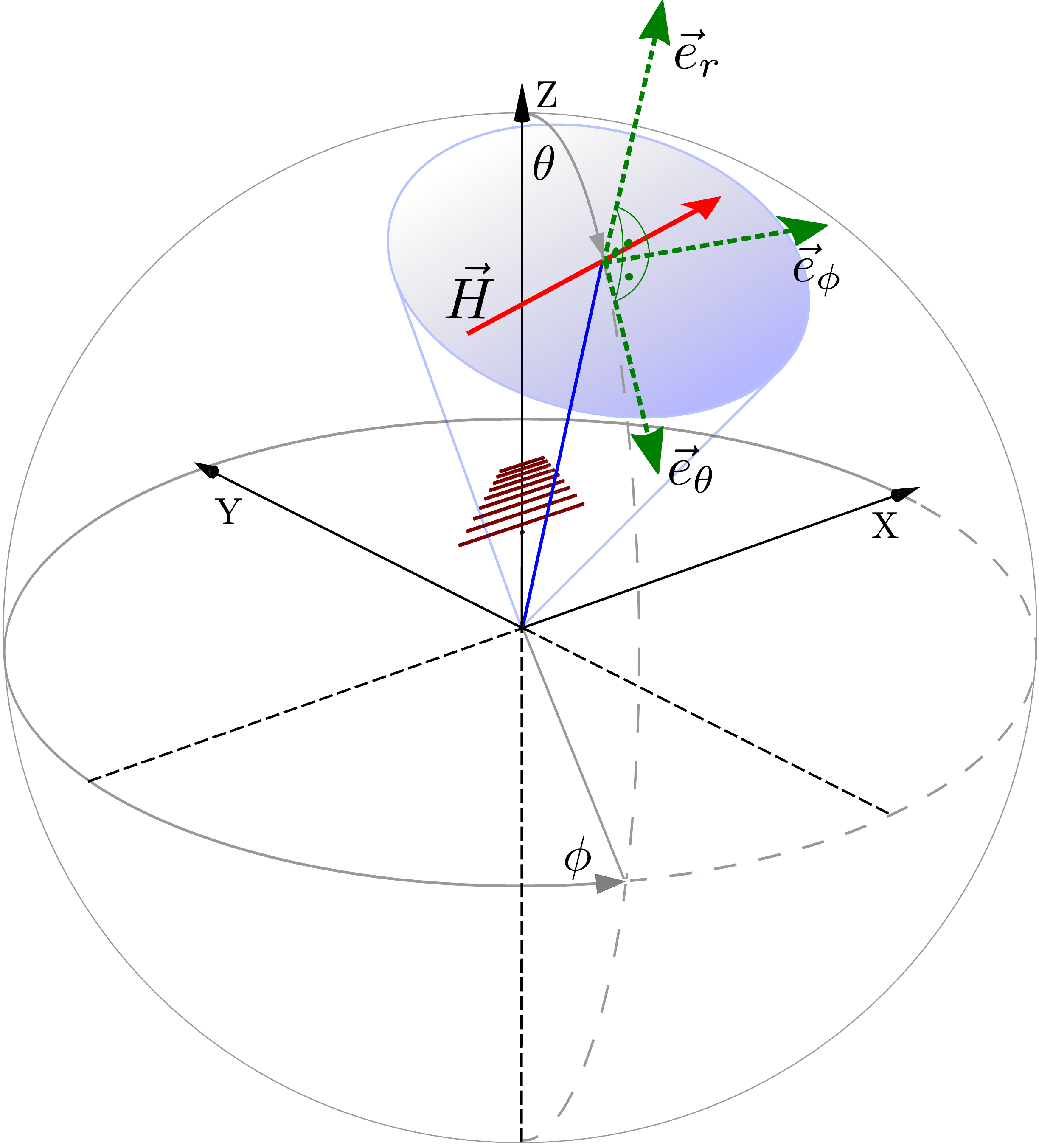}
  \caption{The spherical coordinate system with the antenna structure in the center. The origin of coordinate system is located in the XY-plane below the antenna which takes the placement of the antenna above a ground plane into account. Depicted is a logarithmic periodic antenna structure. The zenith angle $\theta$ is counted from the top, the azimuth angle $\phi$ counterclockwise from the x-axis of the coordinate system. A specific direction ($\theta,\phi$) is considered as the incoming direction of a signal. The vector of the effective antenna length $\vec{H}$ for the specified direction is given. $\vec{H}$ as well as the vector of the electric field (not depicted) are contained in the plane spanned by the unity vectors $\vec{e}_{\theta}$ and $\vec{e}_{\phi}$.}
  \label{fig:VectorHeight}
\end{figure}
For antenna calculations it is convenient to choose a spherical coordinate system with the antenna in its center as depicted in Fig. \ref{fig:VectorHeight}. In this coordinate system the electric field of a plane wave that arrives from a given direction $(\theta,\phi)$ at the antenna will be contained in the plane spanned by the unity vectors $\vec{e}_{\theta}$ and $\vec{e}_{\phi}$ only. The electric field can be written as a two-component vector and is called the instantaneous electric field. Its two components denote two independent polarization directions which vary as a function of time:
\begin{equation}
  \label{eqn:fieldvec}
  \vec{E}(t) =
    \vec{e}_{\theta} \, E_{\theta}(t) + \vec{e}_{\phi} \, E_{\phi}(t) \quad .
\end{equation}

The mapping between the voltage response $V(t)$ induced over the antenna terminals and the electric field $\vec{E}(t)$ is represented by the vector effective length (VEL) $\vec{H}(t)$ of the antenna \cite{Anderson2005}. As displayed in Fig. \ref{fig:VectorHeight} the VEL is a two-component vector in the antenna-based coordinate system as well:
\begin{equation}
  \label{eqn:antennavec}
  \vec{H}(t) =
    \vec{e}_{\theta} \, H_{\theta}(t) + \vec{e}_{\phi} \, H_{\phi}(t) \quad .
\end{equation}
Here, $H_{\theta}(t)$ encodes the response characteristics of the antenna to the component of the incident field in $\vec{e}_{\theta}$-direction and $H_{\phi}(t)$ accordingly. The VEL contains the full information on the response of an arbitrary antenna structure to an arbitrary plane wave signal.

The response voltage of the antenna to a single polarization direction of the electric field is obtained by the convolution of the field and the VEL. For instance in the $\vec{e}_{\phi}$-direction the response is calculated as:
\begin{equation}
 \label{eqn:convolution}
  V_{\phi}(t) = H_{\phi}(t) \ast E_{\phi}(t) \quad ,
\end{equation}
where the symbol '$*$' marks the convolution transform. The total antenna response is obtained as superposition of the partial response voltages $V_{\phi}$ and $V_{\theta}$ to the two independent polarizations of the electric field \cite{Thumm1997}. Using a vectorial form we can write conveniently:
\begin{equation}
  \label{eqn:convolutionvec}
  V(t) = \vec{H}(t) \ast \vec{E}(t) \quad .
\end{equation}
Up to now we have treated the antenna response calculation in the time domain. However, the antenna characteristics contained in $\vec{H}(t)$ are usually accessed in the frequency domain rather than the time domain. We define the vectorial Fourier transforms of the quantities, e.g.:
\begin{equation}
  \vec{\mathcal H}(\omega) \equiv \vec{e}_{\theta} \, \mathcal F(H_{\theta}(t)) + \vec{e}_{\phi} \, \mathcal F(H_{\phi}(t)) \quad .
\end{equation}
and $\mathcal V$, $\vec{\mathcal E}$ accordingly. The script letters indicate complex functions of the angular frequency $\omega = 2\pi\nu$ belonging to the frequency $\nu$.

The convolution theorem 
allows the convolution of functions to be performed
as a point wise multiplication of their Fourier transforms. Hence, the voltage response of Eq. \ref{eqn:convolutionvec} can be treated in the frequency domain as follows:
\begin{equation}
  \label{eqn:response}
  \mathcal V(\omega) = \vec{\mathcal H}(\omega) \cdot \vec{\mathcal E}(\omega) \quad .
\end{equation}
The voltage response in the time domain follows from the inverse Fourier transform $V(t) = \mathcal F^{-1}(\mathcal V(\omega))$. It should be noted that Eq. \ref{eqn:response} represents a condensed way to calculate the antenna response to arbitrary waveforms.

\subsection{Polarization}
\label{sec:Polarization}
\begin{figure}  
  \centering
  \includegraphics[width=\figwidthS\textwidth]{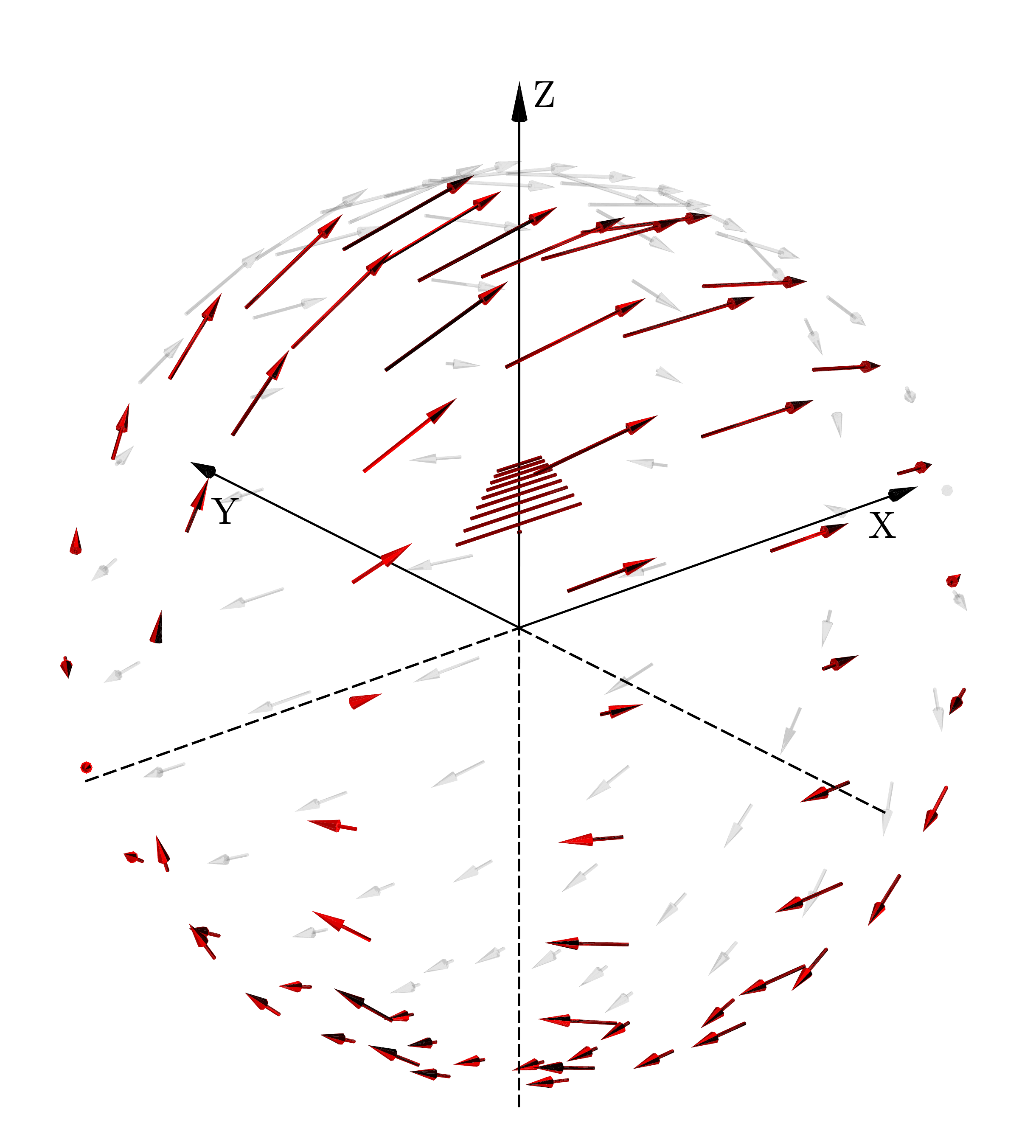}
  \caption{The VEL in the case of the depicted logarithmic periodic dipole structure derived from simulations with the numerical antenna simulation program NEC-2 \cite{Nec} at 75 MHz. For a single frequency the VEL can be expressed as a vector field in the spherical coordinate system of the antenna.}
  \label{fig:VectorField}
\end{figure}

In Fig. \ref{fig:VectorHeight} the antenna-based spherical coordinate system along with an exemplary antenna structure is shown. The pictured vector $\vec{H}$ can either be understood as the VEL at a certain point in time $\vec{H}(t = t_i)$, or at a given frequency $\vec{\mathcal H}(\omega = \omega_j)$. In the latter case the components of $\vec{\mathcal H}$ are complex functions:
\begin{eqnarray}
  \label{ref:jonesvec}
%
%
  \vec{\mathcal H} 
    &=& \vec{e}_{\theta}\,\mathcal H_{\theta} + \vec{e}_{\phi}\,\mathcal H_{\phi} \quad , \\
    &=& \vec{e}_{\theta} \, |\mathcal H_{\theta}| \, e^{i\varphi_{\mathcal H_{\theta}}} + \vec{e}_{\phi} \, |\mathcal H_{\phi}| \, e^{i\varphi_{\mathcal H_{\phi}}} \quad . \label{ref:jonesvec1}
\end{eqnarray}
At a specific frequency the VEL in Eq. \ref{ref:jonesvec} resembles a Jones vector \cite{JONES1941} which is commonly used to describe the polarization of light. Of special interest is the phase difference between the two components:
\begin{equation}
  \Delta \varphi_{\mathcal H} = \varphi_{\mathcal H_{\phi}} - \varphi_{\mathcal H_{\theta}} \, .
\end{equation}
Using this phase difference we separate Eq. \ref{ref:jonesvec1} into a global and a relative phase:
\begin{eqnarray}
  \label{ref:jonesvec2}
  \vec{\mathcal H} 
    &=& e^{i\varphi_{\mathcal H_{\theta}}} \, (\vec{e}_{\theta} \, |\mathcal H_{\theta}| + \vec{e}_{\phi}\,|\mathcal H_{\phi}|\, e^{i\Delta\varphi_{\mathcal H}}) \quad .
\end{eqnarray}
If the phase difference $\Delta \varphi_{\mathcal H}$ at a given frequency $\omega$ is a multiple of $\pi$ 
\begin{equation}
  \Delta \varphi_{\mathcal H} = n\pi\, , \quad  n=\dots,-1,0,1,\dots
\end{equation}
the maximum sensitivity of the antenna is reached for the reception of a linear polarized signal. In this case an intuitive picture of the antenna can be drawn. Following the Jones calculus, we rewrite Eq. \ref{ref:jonesvec2} omitting the global phase:
\begin{equation}
  \label{eqn:VELred}
  \vec{\mathcal H'} = |\vec{\mathcal H}|\,(\vec{e}_{\theta}\,\cos{\alpha} + \vec{e}_{\phi}\,\sin{\alpha})  \quad ,
\end{equation}
where
\begin{equation}
  \alpha = (-1)^n \arctan( \frac{|\mathcal H_{\phi}|}{|\mathcal H_{\theta}|}) 
\end{equation}
denotes the angle of the VEL axis in the $\vec{e}_{\theta}$-$\vec{e}_{\phi}$-plane counted counterclockwise from $\vec{e}_{\theta}$. The VEL $\vec{\mathcal H'}$ is pictured in Fig. \ref{fig:VectorHeight} for a single direction on the unit sphere. In Fig. \ref{fig:VectorField} $\vec{\mathcal H'}$ is displayed for a set of incoming directions. 

The displayed characteristics were accessed via simulations for the displayed logarithmic-periodic antenna structure. The length of the vector changes as a function of zenith angle which denotes the directionality of the antenna. The LPDA is most sensitive to the vertical (zenith) direction.

In the setup shown in Fig. \ref{fig:VectorField}, the VEL vanishes when approaching the x-axis. For the given antenna structure, the electric field of a wave incoming along the x-axis will have no components parallel to the dipoles of the antenna and cannot be detected.

For a simple antenna structure as in Fig. \ref{fig:VectorField}, the VEL is aligned with the projection of the antenna dipole on the unit sphere for a given direction. For any incoming direction of the wave a configuration of the electric field and the effective antenna length exists such that no signal is detected. This is referred to as polarization mismatch.

For single frequencies, the VEL is thus a vector field of 2-dimensional complex vectors in the antenna-based spherical coordinate system. The VEL thus has three major dependencies:
\begin{equation}
  \vec{\mathcal H}(\omega) = \vec{\mathcal H}(\omega,\theta,\phi) \quad .
\end{equation}

In the general case of an elliptical polarization where $\Delta \varphi_{\mathcal H} \neq n\pi$ the VEL can be presented as an ellipse on the unit sphere rather than a vector. However, in the case of the simulated antenna, the VEL is a vector.
The omission of a global phase in Eq. \ref{eqn:VELred} enables the inspection of the polarization of an antenna at single frequencies. The wideband characteristics of antennas are contained in the development of $\vec{\mathcal H}$ as a function of frequency.

The VEL is related to the antenna gain and the antenna directivity. In Appendix \ref{app:VELandGain} we briefly discuss its relationship to these more commonly used quantities.   

In the case of the antenna models discussed in Sec. \ref{sec:Antennas}, two rotated antennas are assembled with the same hardware structure. In Appendix \ref{app:EFieldReco} we show how the VEL can be applied to reconstruct the 3-dimensional electric field vector of the signal recorded in such antenna setups. In Appendix \ref{app:EFieldReco} we also relate the VEL to the Jones antenna matrix used in radio polarimetry \cite{J.P.Hamaker1996a}.

\subsection{The Realized Vector Effective Length in a Measurement Setup}
\label{sec:RVEH}
The VEL as defined in Sec. \ref{sec:VectorEffectiveLength} relates the incident electric field to an open circuit voltage at the antenna terminals which we will refer to as $\mathcal V_{\mbox{\small oc}}$ in the following.
In an actual measurement setup the antenna will be read out at a load impedance. In this section we will focus on the impact of the readout system on the measured signal.

\subsubsection{The Thevenin Equivalent Antenna}
\label{sec:Thevenin}
\begin{figure*}
  \begin{minipage}[c]{0.50\textwidth}
    \centering
    \includegraphics[width=.6\textwidth]{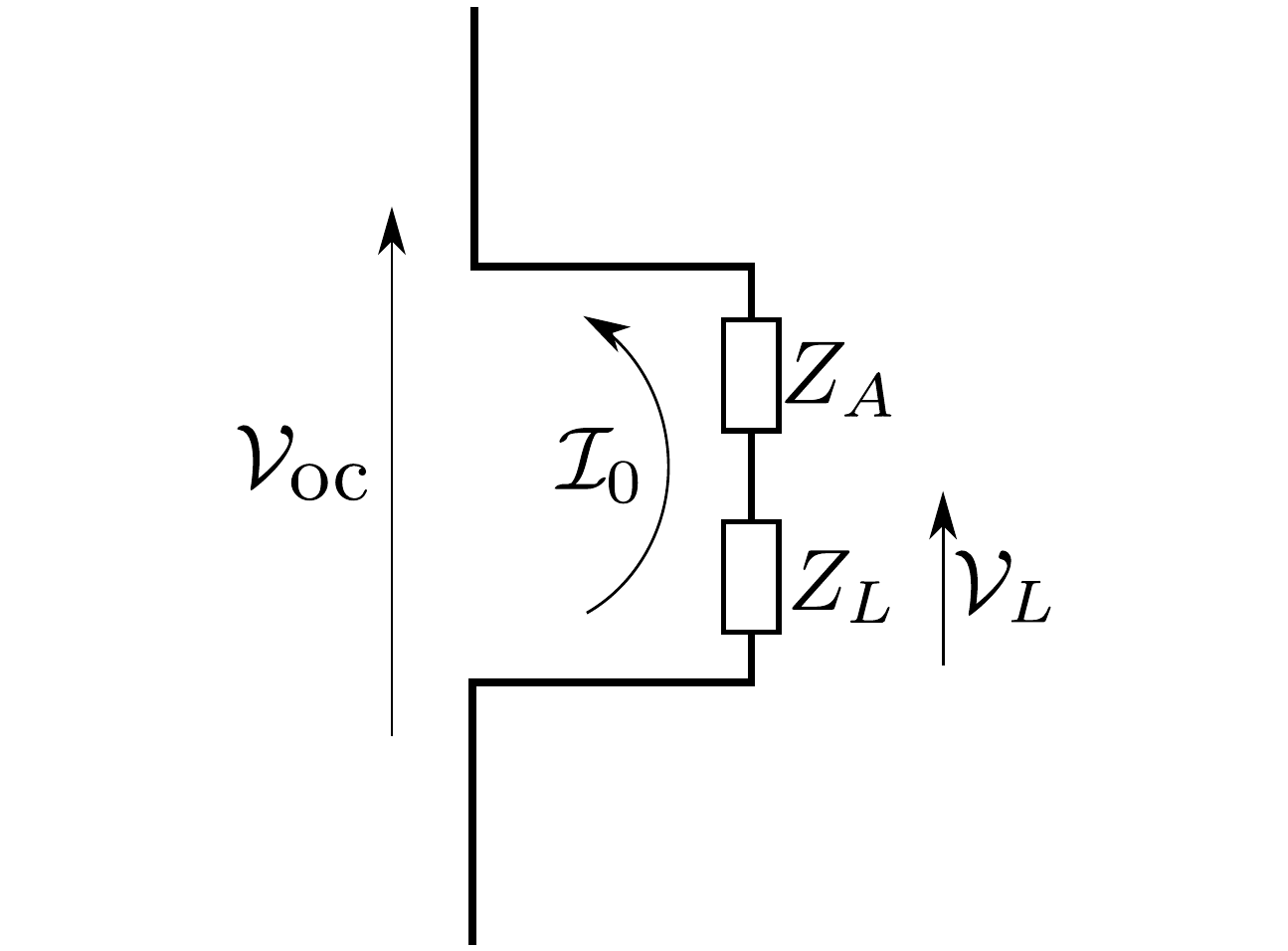}
  \end{minipage}
  \begin{minipage}{0.50\textwidth}
    \centering
    \includegraphics[width=.6\textwidth]{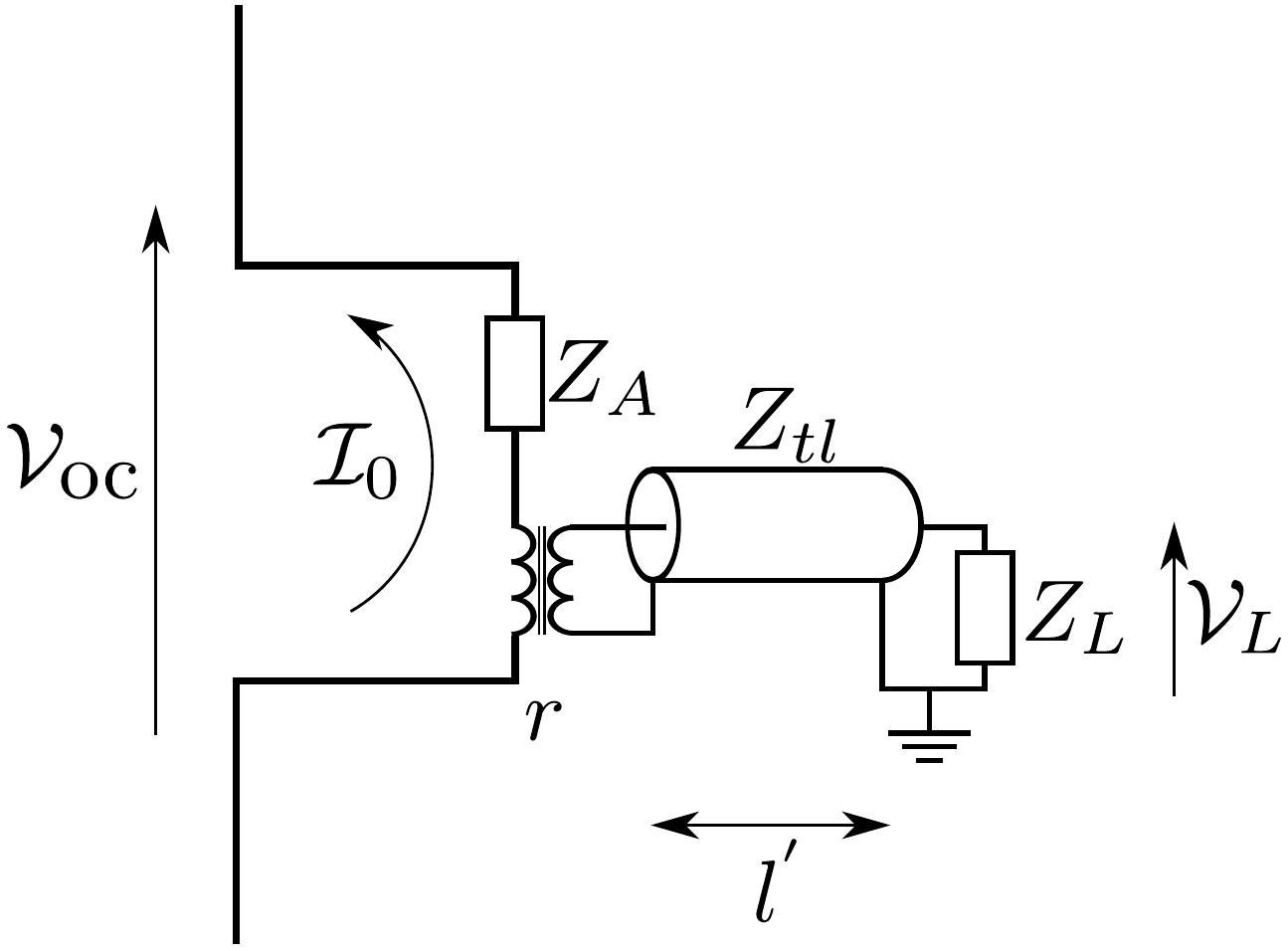}
  \end{minipage}
  \caption{Left: Thevenin equivalent of an active antenna used for reception. Right: Thevenin equivalent of an antenna used for reception with intermediate transmission line and transmission line transformer. Please refer to the text for a description of the symbols.}
  \label{fig:Cir}
\end{figure*}
In Fig. \ref{fig:Cir} (left) the Thevenin equivalent circuit diagram for a simple measurement situation is displayed. The antenna is read out introducing a load impedance $Z_L$ in addition to the antenna impedance $Z_A$. The voltage measured over the load impedance follows from the voltage divider relation:
\begin{equation}
 \label{eqn:simpleMatch}
 \mathcal V_L = \frac{Z_L}{Z_A+Z_L} \mathcal V_{\mbox{\small oc}} \equiv \rho \, \mathcal V_{\mbox{\small oc}} \quad .
\end{equation}
The impedances $Z_A$ and $Z_L$ are complex functions of the frequency $\omega$. The situation of conjugate matching (cf. Appendix \ref{app:VELandGain}) is obtained when the readout impedance is the complex conjugate of the antenna impedance $Z_L = Z_A^*$. 

If the complex transfer function $\rho$ in Eq. \ref{eqn:simpleMatch} is included in the formulation of the measurement equation (Eq. \ref{eqn:response}), the VEL is referred to as the realized or normalized VEL $\vec{\mathcal H}_r$ \cite{Soergel2005}:
\begin{equation}
 \label{eqn:realizedVEL}
 \mathcal V_L = \rho \, \vec{\mathcal H} \cdot \vec{\mathcal E} = \vec{\mathcal H}_r \cdot \vec{\mathcal E} \quad .
\end{equation}
The transfer factor $\rho$ as given in Eq. \ref{eqn:simpleMatch} is sufficient to calculate the response voltage in the case of the Butterfly antenna (cf. Sec. \ref{sec:Antennas}). For the other antennas, more elaborate diagrams are required.

\subsubsection{Transformers for Impedance Matching}
\label{sec:Transformer}
The transfer function $\rho$ may include more complex setups than the one displayed in Fig. \ref{fig:Cir} (left). The small aperiodic loop antenna Salla includes a transformer to realize a better matching between the antenna impedance $Z_A$ and the impedance of the readout amplifier $Z_L$. 

We denote the impedance transformation ratio of the transformer by $r$. In the circuit diagram shown in Fig. \ref{fig:Cir} (left) the load impedance due to the combination of readout impedance and transformer is then $\,r \cdot Z_L\,$. Along with impedance transformation the transformer changes the voltage that is delivered to the readout impedance. Here an additional factor $1/\sqrt{r}$ needs to be added \cite{Davis2001}. Hence, if an ideal transformer is used to optimize matching, the transfer factor $\rho$ becomes:
\begin{equation}
  \label{eqn:Transformer}
  \rho = \frac{1}{\sqrt{r}}\frac{r\, Z_L}{Z_A + r Z_L} \quad .
\end{equation}

\subsubsection{Intermediate Transmission Lines}
\label{sec:ITL}
The calculations performed in Sec. \ref{sec:Thevenin} and \ref{sec:Transformer} implicitly assumed that electric distances between the position of the impedances are short in comparison to the wavelength processed. However, the Small Black Spider LPDA uses a transformer to feed the antenna signals into a coaxial cable which guides them to the first amplifier. With increasing length of the coaxial cable, the direct current approximation becomes invalid and propagation effects need to be taken into account. A circuit diagram for this setup is displayed in Fig. \ref{fig:Cir} (right).
In Appendix \ref{app:IntermediateTransmissionLines} we derive a transfer factor $\rho$ that enables calculations of multiple signal reflections between readout impedance and antenna to be made as a single step in the frequency domain. It is given by:
\begin{equation}
  \label{eqn:realizedVELtotal}
  \rho = \frac{\sqrt{r}\,Z_{tl}}{Z_A + r\,Z_{tl}} \cdot  \frac{(1+\Gamma_L)\,e^{\gamma l'}}{e^{\gamma 2l'}-\Gamma_A\, \Gamma_L} \, .
\end{equation}
Here, $Z_{tl}$ is the characteristic impedance of the intermediate transmission line and $l'$ its electrical length. The complex propagation constant $\gamma$ per unit electrical length includes the attenuation loss along the transmission line. $\Gamma_L$ and $\Gamma_A$ are the voltage reflection coefficients from the transmission line to the load and from the transmission line to the antenna respectively.
 
The results for the less complex setups discussed in the previous subsections are included in Eq. \ref{eqn:realizedVELtotal}, e.g. when the transmission line is short: $l' \rightarrow 0$ or the transformer ratio is $r=1$. Hence, Eq. \ref{eqn:realizedVELtotal} unifies the calculation for the three antennas discussed in Sec. \ref{sec:Antennas}.

In the case of the Small Black Spider LPDA, the electrical line length is $l'\approx 4.4 \,$m. With respect to the discussion in this section special care was taken during the design of the amplifier impedance to match the $50\, \Omega$ transmission line. Hence, Eq. \ref{eqn:realizedVELtotal} introduces only slight changes to the signal shape and a time delay due to the length of the transmission line.
\begin{figure*}
  \begin{minipage}[c]{0.33\textwidth}
    \centering
    \includegraphics[width=0.8\textwidth]{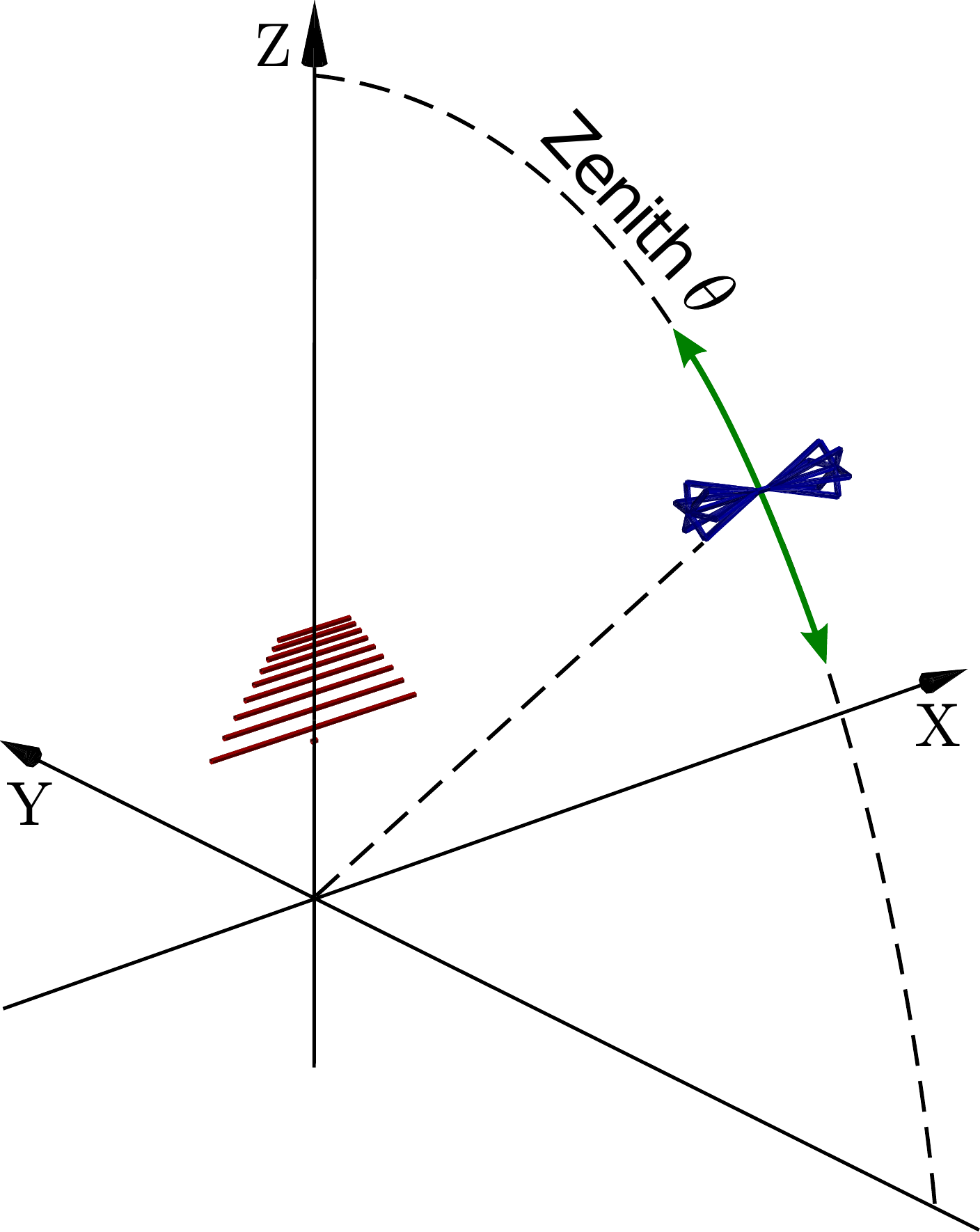}
  \end{minipage}
  \begin{minipage}{0.33\textwidth}
    \centering
    \includegraphics[width=0.8\textwidth]{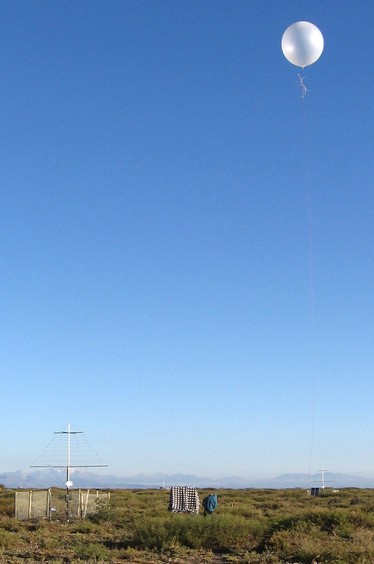}
  \end{minipage}
  \begin{minipage}{0.33\textwidth}
    \centering
    \includegraphics[width=0.8\textwidth]{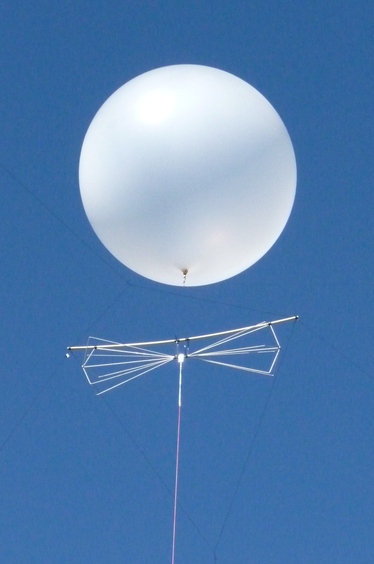}
  \end{minipage}
    \caption{The setup for calibration measurements of the Small Black Spider LPDA at the site of AERA. Left: Sketch of the measurement setup. A calibrated transmitter antenna is moved on a circle around the antenna under test to access different zenith angles. Only the read out plane of the tested antenna is displayed. Middle: A picture of the setup. In the lower left the AERA station used for the calibration measurement is visible. At a distance of $\sim$30 m a balloon carries the calibrated transmitter antenna. The position of the transmitter is fixed by a three-legged rope system which provides the movement on the circle and a parallel orientation of the transmitter and receiver antenna. Right: Picture of the calibrated biconical antenna used as transmitter. The biconus has a length of 1.94 m. The antenna is carried by a balloon filled with 5 m$^3$ helium and is fed by a coaxial cable running to the ground. The overall weight lifted by the balloon is $\sim$4.5 kg.}
    \label{fig:CalibrationSetup}
\end{figure*}

\subsection{Signal Amplification}
\label{sec:SignalAmplification}
With Eq. \ref{eqn:realizedVEL} the incident electrical field is related to the voltage $\mathcal V_L$ over the readout impedance of the antenna. For the antennas being considered this readout impedance is the input impedance of an LNA. Amplifiers are characterized by the complex scattering or S-parameters \cite{Carlin1956}. The S-parameter $\mbox{S}21$ is the ratio of the amplified to the incoming voltage amplitude. It should be noted that the amplification characteristics expressed by $\mbox{S}21$ implicitly assume that the amplifier is operated within a system with the same impedance as during the S-parameter measurement. Looking e.g. at Fig. \ref{fig:Cir} (left) this assumption is not fulfilled: the antenna impedance $Z_A$ replaces the generator impedance used during the S-parameter measurement and may itself be frequency dependent. In Appendix \ref{app:Sparameter} we show that, in this case, a corrected amplification factor $\mbox{S}21'$ --- also referred to as voltage gain --- can be used.
 This is 
normalized to the voltage over the amplifier input impedance $\mathcal V_L$  during the S-parameter measurement:
\begin{equation}
  \label{eqn:sParameter}
  \mbox{S}21' = \frac{\mbox{S}21}{1+\mbox{S}11} \quad .
\end{equation}
Here, $\mbox{S}11$ corresponds to the voltage reflection coefficient at the input of the amplifier during the S-parameter measurement.
In the case of the antennas discussed, $\mbox{S}21'$ relates the voltage $\mathcal V_{a}$ at the amplifier to the voltage amplitude fed into a well defined $50 \, \Omega$ system:
\begin{equation}
  \label{eqn:Vamp}
  \mathcal V_{a} = \mbox{S}21' \cdot \mathcal V_L \quad .
\end{equation}
This amplified voltage is used for further signal processing. $\mathcal V_{a}$ is of special interest as signal to noise ratios are essentially fixed in a readout chain after the first amplification. The noise performance is not discussed here but will be determined experimentally in Sec. \ref{sec:ObsGal} for the discussed antennas.

Especially for antennas where the LNA is directly integrated into the antenna structure, e.g. for the Butterfly antenna, it is useful to define a VEL that includes the amplification of the signal:
\begin{equation}
  \label{eqn:VELamp}
  \vec{\mathcal H}_a \equiv \mbox{S}21' \cdot \vec{\mathcal H}_r \quad .
\end{equation}
$\vec{\mathcal H}_a$ is convenient to access in antenna calibration measurements as we will show in Sec. \ref{sec:CalMeasurement}. Using Eqs. \ref{eqn:Vamp} and \ref{eqn:realizedVELtotal} allows us to take the impact of the amplifier and the readout system into account when studying antenna characteristics simulated with the NEC-2 program in the next sections.

  \section{Measurement of the Characteristics of the Small Black Spider LPDA}
\label{sec:CalMeasurement}
In Sec. \ref{sec:AntennaTheory} we have recognized that the vector effective length is the central antenna characteristic needed to perform  calibrated measurements of the electric field strength of incoming radio signals. The VEL can be calculated in simulations. However, it is desirable to perform the unfolding of the impact of the detector on the recorded signal on the basis of measured antenna properties. 

In this section we present calibration measurements we have performed to access the VEL of a Small Black Spider LPDA installed as part of the first stage of AERA. In our measurements we focus on the zenith angle dependence of the VEL as a function of frequency. In preceding measurements we have confirmed that the azimuthal dependency of the VEL follows a simple sinusoidal function as expected for dipole-like antennas such as the Small Black Spider LPDA \cite{Seeger2010a}.

\subsection{Antenna Calibration Setup}
\label{sec:CalSim}
So far, directional properties of antennas for the detection of cosmic rays have been measured with down-scaled versions of the antenna under test \cite{Kroemer2008} or have been constrained to a measurement of the scalar amplitude transfer without the vectorial phase information \cite{Nehls2008}. 

The experimental challenge in a calibration measurement with a full scale radio antenna is to fix a distance between transmitter and receiver antenna such that the wave emitted by the transmitter sufficiently approximates the plane wave condition at the tested antenna.
It should be noted that the usual approximation at which distance $r_{\mbox{\small ff}}$ this far field condition is fulfilled
is given by $r_{\mbox{\small ff}} > 2\,\delta^2/\lambda$, where $\delta$ the largest dimension of the transmitting antenna. This relation is valid only if $\delta$ is \textit{larger} than the wavelength $\lambda$ \cite{Balanis2005}. This is typically the case for dish antennas but not for the radio antennas discussed in this article. 

With respect to the longest wavelength of the AERA bandwidth (10 m), distances between transmitter and the tested antenna of 3$\lambda$ are realized in our calibration. Following the discussion in Ref. \cite{Kraus2003} we can estimate the impact of near-field components still present in the calibration measurement. At a distance of $3\lambda$ these will cause a variation of the power angular distribution of at most $\pm0.5$ dB when compared to a measurement performed at much larger distances.

In Fig. \ref{fig:CalibrationSetup} an overview of the calibration setup is given. To access large distances at small zenith angles above the antenna under test, a balloon is used to lift a calibrated transmitter antenna \cite{Schwarzbeck}.
Ropes constrain the movement of the transmitter to a circle around the tested antenna and ensure a parallel orientation of the two antennas. In this way the $\vec{e}_{\phi}$-component $\mathcal H_{\phi}$ of the VEL can be accessed for various zenith angles.

The transmission measurement from the biconical antenna to the Small Black Spider is performed using a vector network analyzer \cite{FSH4}.
The network analyzer simultaneously feeds the transmitter antenna and reads out the tested antenna. The signal delivered by the vector analyzer is adjusted to appear $>$ 30 dB above the ambient radio background recorded by the tested antenna throughout the measurement bandwidth from 30 to 80 MHz. The impact of the coaxial cables needed for the connections are removed from the data by including them in the null calibration of the vector network analyzer prior to the measurement. The amplifier of the Small Black Spider is included in the transmission measurement. Hence, the setup allows us to measure $\mathcal H_{a,\phi}(\omega,\theta,\phi=270^{\circ})$ as discussed in Eq. \ref{eqn:VELamp} of Sec. \ref{sec:SignalAmplification}.

\subsection{Simulated Calibration Setup}
To cross-check the calibration measurement procedure we performed simulations of an equivalent setup using the numerical antenna simulation tool NEC-2 \cite{Nec}. The simulated calibration setup includes a model of the Small Black Spider LPDA as well as a model of the biconical antenna used as a transmitter. Both antenna models are placed in the simulation according to the geometries existing in the actual calibration setup. The sketch in Fig. \ref{fig:CalibrationSetup} (left) is generated from 
a NEC-2 simulation of a transmission measurement.

The simulation model of the transmitter antenna is excited by placing a voltage source at its footpoint. The NEC-2 simulation then calculates the power emitted by the transmitter and the consequent open terminal voltage  $\mathcal V_{\mbox{\small oc}}$ induced in the structure of the Small Black Spider. Following the discussion in Sec. \ref{sec:RVEH} we process the terminal voltage to give the power delivered into a $Z_{tl}=50\,\Omega$ system which corresponds to the coaxial cables connected to the network analyzer. 

In the field measurements, not all power that is delivered by the signal source of the network analyzer is accepted by the transmitter due to impedance mismatch. This has to be accounted for in the simulation by dividing the radiated power calculated by NEC-2 with the measured power acceptance of the transmitter antenna.

In the measurements at the AERA site, signal reflections from the ground are included. We thus use the option of NEC-2 to model a ground plane in the simulation using the Fresnel reflection coefficients. The reflection coefficients depend on the relative permittivity $\varepsilon_r$ and the conductivity $\sigma$ of the soil. With $\sigma = 0.0014\,\,\Omega^{-1}\mbox{m}^{-1}$ we assume a low conductivity which has been confirmed in initial test measurements at the AERA site. For low conductivities we find $\varepsilon_r = 5.5$ to be a reasonable choice for the relative permittivity in typical ground scenarios \cite{ITURecommendation1992}. The resulting ground exhibits a relatively low reflectivity.

As in real measurements, we use the simulation to yield the ratio of signal received by the tested antenna to the signal used to operate the transmitter. Simulated and measured data are then processed equally in further analysis.

\subsection{Transmission Equation and Data Processing}
For each zenith angle accessed in the setup we measure the S-Parameter S21 as a function of frequency. S21 is the complex ratio of the voltage amplitude $\mathcal V_a$ delivered by the tested antenna and the amplitude $\mathcal V_g$ delivered from the signal generator to the transmitter:
\begin{equation}
  \label{eqn:s21}
  \mbox{S}21 = \frac{\mathcal V_a}{\mathcal V_g} \quad .
\end{equation}
The voltage $\mathcal V_a$ is the response of the tested antenna to the electric field $\mathcal E_{\phi}^t$ caused by the transmitter antenna:
\begin{equation}
  \label{eqn:V_a}
  \mathcal V_a = \mathcal H_{a,\phi} \, \mathcal E_{\phi}^t \quad .
\end{equation}
Due to the configuration of the setup, the electric field is contained in the $\vec{e}_{\phi}$-direction of the antenna-based coordinate system. In antenna theory (e.g. Ref. \cite{Balanis2005}) the electric field of a transmitting antenna at a distance R is given. For our calibration setup it is:
\begin{equation}
  \label{eqn:EField}
  \mathcal E_{\phi}^t = -i \,\, Z_0 \,\, \frac{1}{2\,\lambda\, R} \, \mathcal I_0^t \,\, \mathcal H_{\phi}^t \,\, e^{-i\,\omega\,R/c} \quad ,
\end{equation}
where $H_{\phi}^t$ is the VEL, $\mathcal I_0^t$ denotes the current in the transmitter antenna as depicted in Fig. \ref{fig:Cir}, and $c$ is the speed of light. When Eqs. \ref{eqn:V_a} and \ref{eqn:EField} are inserted in Eq. \ref{eqn:s21} we obtain a complex form of the Friis transmission equation:
\begin{equation}
  \label{eqn:Friis}
  \mbox{S}21 = -i \, Z_0 \, \frac{1}{2\,\lambda\, R} \, \frac{\mathcal I_0^t}{\mathcal V_g} \,\, \mathcal H_{\phi}^t \,\, \mathcal H_{a,\phi} \,\, e^{-i\,\omega\,R/c} \quad .
\end{equation}
To access the desired VEL $\mathcal H_{a,\phi}$ of the antenna under test, the characteristics of the calibrated transmitting antenna $\mathcal H_{\phi}^t$ have to be applied. In our case these are given in terms of realized gain $G_r$. The realized gain refers to the transfer of signal power rather than signal amplitude and includes the reflections at the input of the transmitter antenna. In Appendix \ref{app:velandrelgain} we derive that:
\begin{equation}
  \label{eqn:app1}
 \frac{|\mathcal I_0^t|}{|\mathcal V_g|} \, |\mathcal H_{\phi}^t|  = 
 \sqrt{\frac{\lambda^2}{\pi \, Z_0 \, Z_{tl}}\,G_{\mbox{\footnotesize cal}}^t} \quad .
\end{equation}
The transmitter calibration $G_{\mbox{\footnotesize cal}}^t(\omega)$ is given by the manufacturer of the antenna as a function of real numbers. 
Such simplification is acceptable if the transmitter antenna introduces only minor distortions to the phasing of the signal within the measurement bandwidth.
This has been verified in preceding test measurements for the biconical antenna \cite{Seeger2010a}.

The combination of Eqs. \ref{eqn:Friis} and \ref{eqn:app1} yields the measurement equation for the calibration setup:
\begin{equation}
  \label{eqn:MEQN}
  \mathcal H_{a,\phi} = i \,\, R \,\, \mbox{S}21 \, \sqrt{\frac{Z_{tl}}{Z_0}} \, \sqrt{\frac{4\pi}{G_{\mbox{\footnotesize cal}}^t}} \, e^{i\,\omega\,R /c } \quad ,
\end{equation}
where we measure the distance R between the center of the transmitting antenna and the center of the lowest dipole of the Small Black Spider LPDA.

\subsection{Calibration Measurement Results}
\label{sec:CalMeasRes}
\begin{figure}
  \centering
  \includegraphics[width=\figwidth\textwidth]{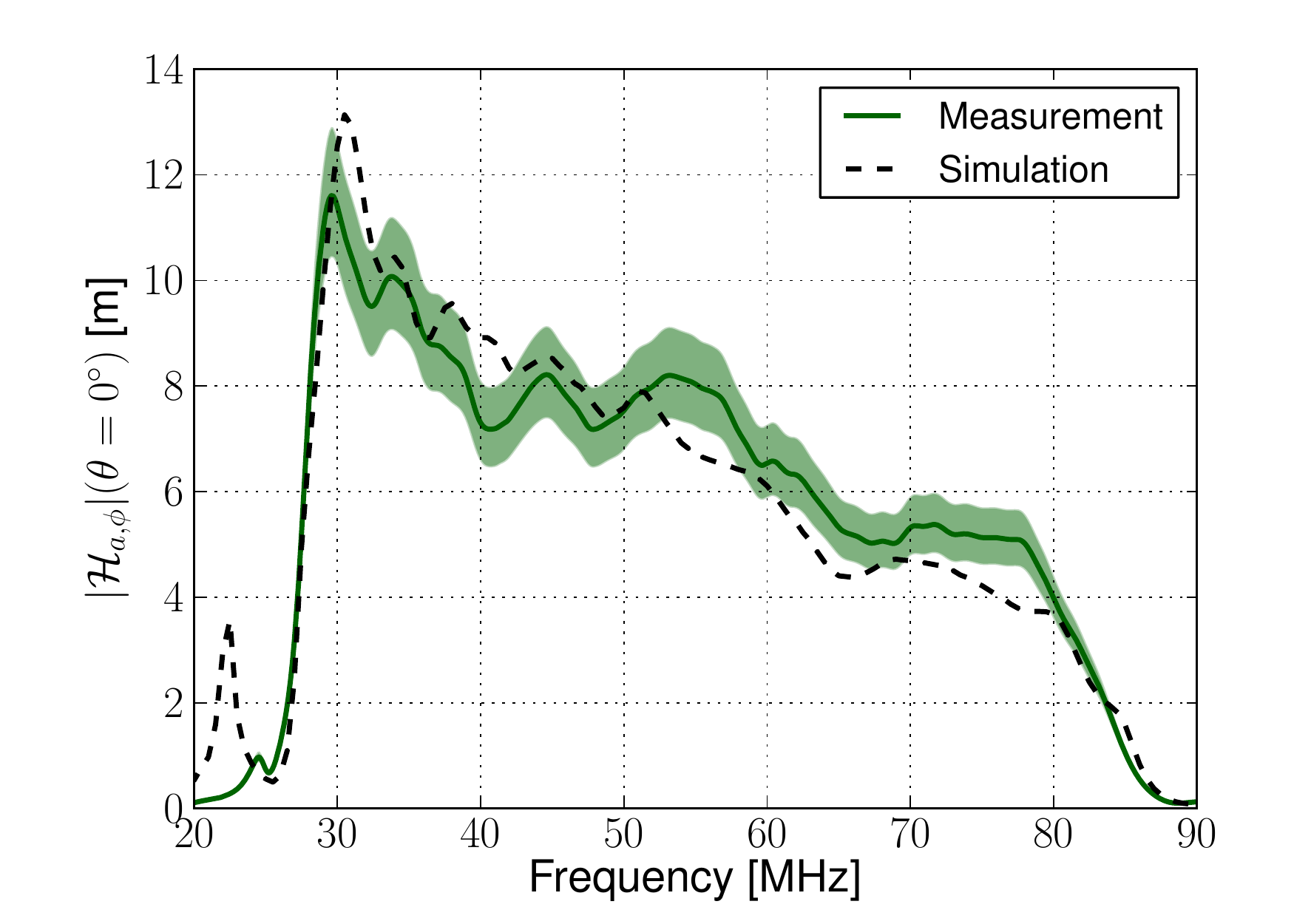}
  \caption{The amplified vector effective length of the Small Black Spider LPDA for the zenith direction as a function of frequency in measurement and simulation. The uncertainty of the measurement is indicated by the shaded area and is dominated by the systematic uncertainties of the calibration of the transmitting antenna of 0.7 dB and by the precision of the transmission measurement, here 0.6 dB.}
  \label{fig:GainMeasured}
\end{figure}
\FIGURE{
  \centering
  \includegraphics[width=1.\textwidth]{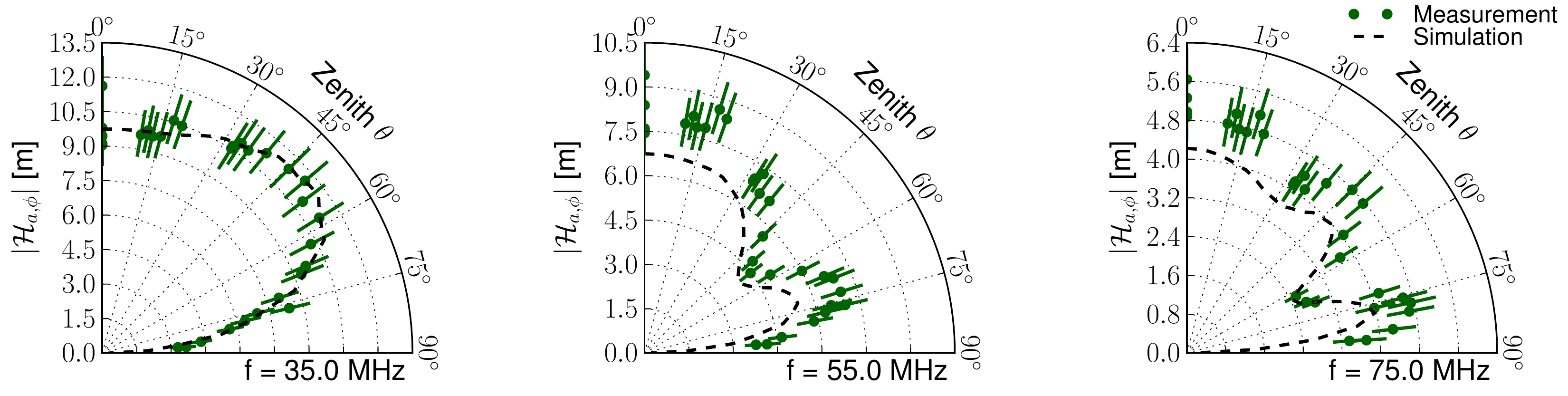}
  \caption{The VEL of the Small Black Spider LPDA as a function of zenith angle for three different frequencies in measurement and simulation. Data points are given for all zenith directions that we were able to access with the setup shown in Fig. \ref{fig:CalibrationSetup} . The uncertainty of the measurement is indicated by the bars analog to Fig. \ref{fig:GainMeasured}.}
  \label{fig:VELvsZen}
}
In Fig. \ref{fig:GainMeasured} the absolute value of the amplified VEL of the Small Black Spider is displayed for the zenith direction $\mbox{$\theta = 0^{\circ}$}$ as a function of frequency. For frequencies lower than 30 MHz and higher than 80 MHz the reception is strongly suppressed due to the presence of the filter elements in the amplifier (cf. Sec. \ref{sec:SBS}). Within the bandwidth the VEL decreases with increasing frequency. This is the expected behavior for antennas which feature a gain that is constant as a function of frequency, such as LPDAs (cf. Eq. \ref{eqn:gain}). The additional variations of the VEL within the frequency band occur due to the interplay of the LPDA's dipole elements which resonate at different frequencies. We observe that the simulation reproduces the bandwidth and the overall size of the measured VEL.

For a set of three frequencies, the dependence of the VEL on the zenith angle is shown in Fig. \ref{fig:VELvsZen}.
For the low frequencies, the antenna is most sensitive to zenith angles around $45^{\circ}$. At higher frequencies, a side-lobe pattern evolves with up to two lobes at the highest frequencies. 

The primary cause for the side lobes is the constructive and destructive interference of the direct wave and the wave reflected from the ground at the position of the antenna with its lowest dipole at a height of 3 m. Note that a conclusion on the reception of transient signals can only be drawn if the wide band combination of these patterns including their respective phasing is regarded, as we will do in Sec. \ref{sec:SimTrans}. 

With respect to the shape of the side lobe pattern we note a good agreement between measurement and simulation. 
For the combination of all zenith directions 
and all frequencies within the bandwidth we observe an agreement of the simulated and measured VEL of better than $\pm 20 \%$.

\begin{figure}
  \centering
  \includegraphics[width=\figwidth\textwidth]{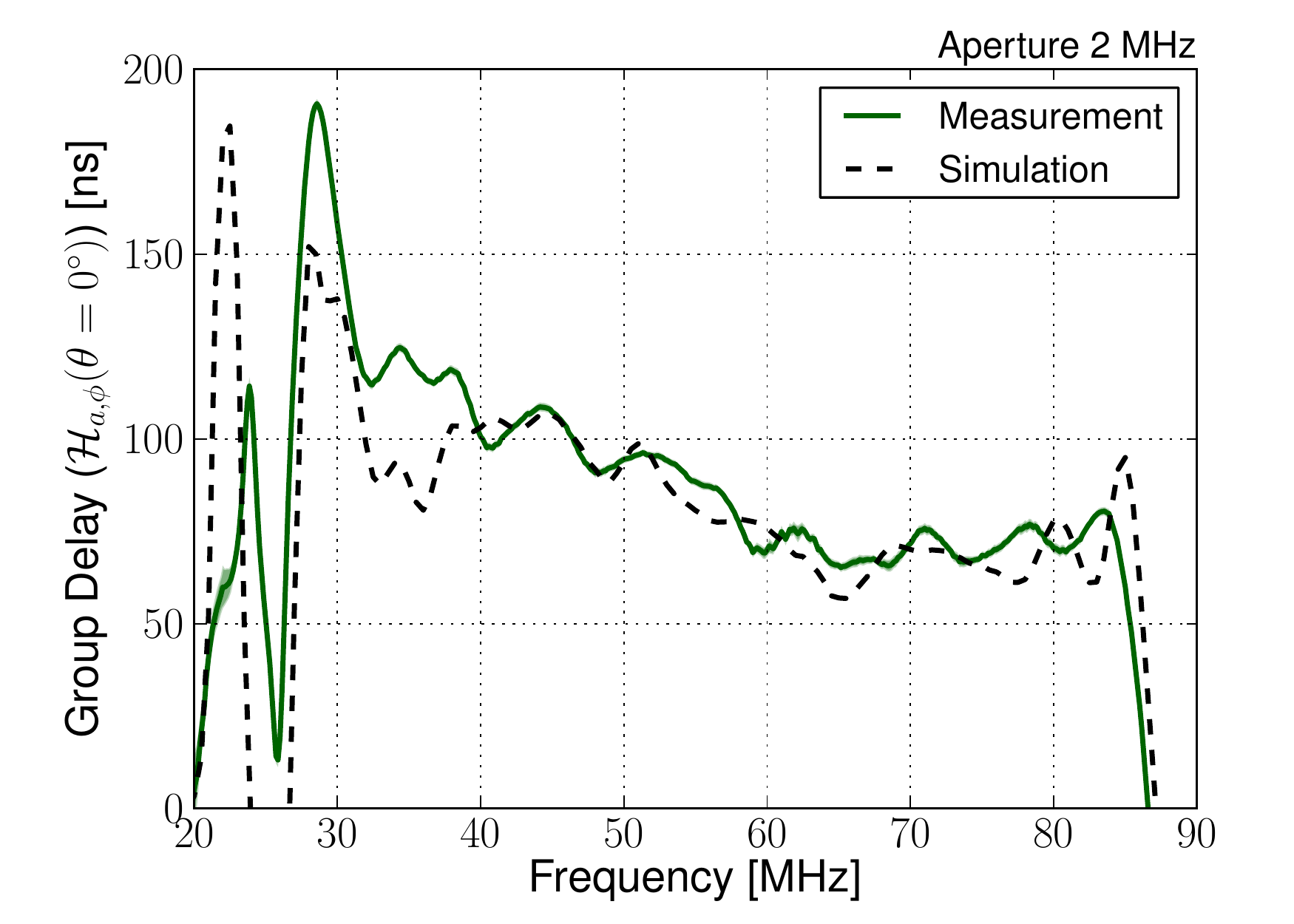}
  \caption{The group delay including the LNA for the zenith direction $\theta=0^{\circ}$ as a function of frequency for the Small Black Spider LPDA in simulation and measurement. The uncertainty of the group delay is $<$ 1.3 ns and results from variations observed in multiple measurements.}
  \label{fig:GDMeasured}
\end{figure}
\begin{figure*}
  \centering
  \includegraphics[width=1\textwidth]{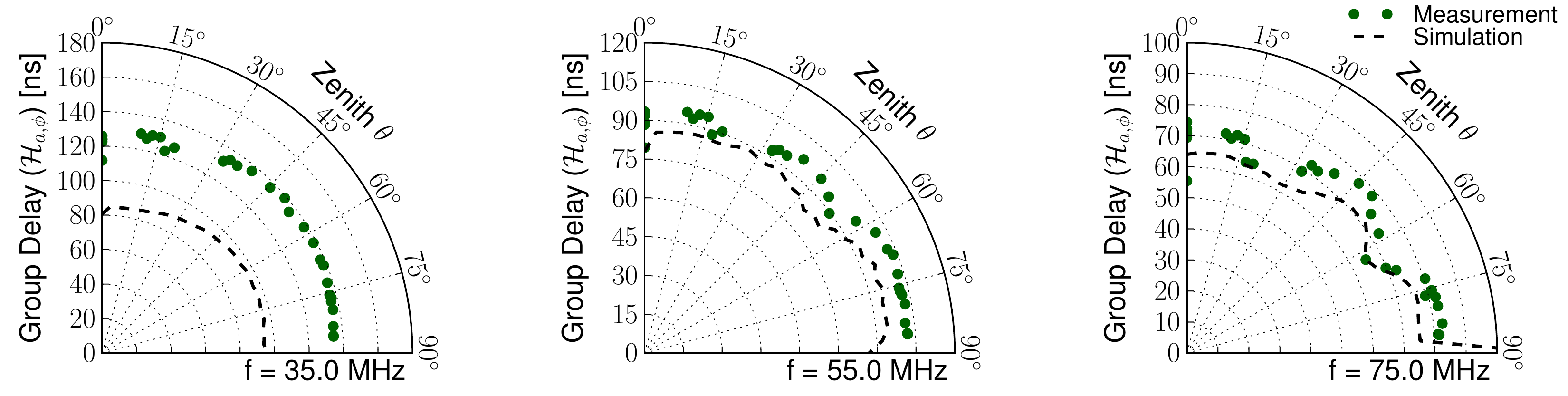}
  \caption{The group delay of the Small Black Spider LPDA as a function of zenith angle for three different frequencies in the measurements and the simulations.}
  \label{fig:GDvsZen}
\end{figure*}

The phasing of $\mathcal H_{a,\phi}$ reveals the group delay induced by the Small Black Spider to the transmitted signal. The group delay $\tau$ is given by:
\begin{equation}
  \label{eqn:goupdelay}
  \tau(\omega) = -\frac{d}{d \omega} \arg(\mathcal H_{a,\phi}) \quad .
\end{equation}
In Fig. \ref{fig:GDMeasured} the group delay for the zenith direction is displayed.
Within the measurement bandwidth from 30 to 80 MHz the group delay decreases by $\sim 50$ ns, where measurement and simulation agree on the functional dependence. 

As will be discussed in detail in Sec. \ref{sec:CompTransResp} a non-constant group delay induces the dispersion of the observed transient signal and thus reduces its peak amplitude. The group delay displayed in Fig. \ref{fig:GDMeasured} results from a combination of the delay introduced by the logarithmic periodic structure of the Small Black Spider and the delay introduced by the filter elements of the amplifier. Although the amplifier has been designed to suppress the signal reception outside the bandwidth, it causes a non constant group delay also within the measurement bandwidth especially at lower frequencies. The most recent version of the AERA readout electronics is able to compensate digitally for the dispersion induced by non constant group delays of the readout chain which we will report in a later article.

In Fig. \ref{fig:GDvsZen} the group delay for the Small Black Spider is shown as a function of zenith angle for three different frequencies. We also find that the group delay exhibits a side-lobe pattern that is similar to the pattern observed in the case of the absolute values of the VEL. The shape of the group delay pattern is similar in the measurements and the simulations. The absolute values differ by up to $30\, \%$ depending on the considered frequency range, as is also visible in Fig. \ref{fig:GDMeasured}.

Correction for a group delay that changes with incoming direction is important as single transient signal fronts will be observed at different detector stations.

In the present calibration campaign we have measured the $\vec{e}_{\phi}$-component of the VEL. The full VEL will be obtained in later calibration measurements.

From the comparison of the measured and the simulated antenna characteristics we conclude that overall the simulations give a realistic description of the Small Black Spider antenna characteristics.

  \section{Comparison of Transient Antenna Responses}
\label{sec:CompTransResp}
In this section we use simulations to compare the three antennas presented in Sec. \ref{sec:Antennas} with respect to their response to transient signals. 

\subsection{Simulation of the Vector Effective Length}
\label{sec:SimTrans}
To investigate the transient response characteristics of the antennas we simulated the respective VELs with the NEC-2 program. In the case of the Small Black Spider, the simulated antenna model is identical to the one presented in Sec. \ref{sec:CalMeasurement}. Simulation models for the Butterfly and the Salla antennas have been created following their corresponding structure specifications. The Butterfly antenna is explicitly designed to be used at a height of 1.5 m above ground, which we have set in the simulation. The other two antennas are simulated with the lowest sensitive element at 3 m, the current installation height of the AERA antennas.

For the simulated calibration measurements in Sec. \ref{sec:CalMeasurement} we have explicitly introduced a transmitting antenna as the source of the electric field. 
For the comparison between the different antennas we access the VEL from direct simulations of the receiving antenna without an additional transmitter. The simulated antenna characteristics presented in this section thus avoid possible deficiencies of the transmitter simulation and would correspond to calibration measurements performed at large distances.

We use the antenna simulation program NEC-2 to access the VEL directly by exciting the tested antenna with a plane wave. Here the simulation computes the resulting currents in the receiving antenna which enables calculations of the response voltage as described in Sec. \ref{sec:CalSim}. For each incoming direction, the two components of the VEL are accessed as the ratio of the response voltage and electric field amplitude by choosing the polarization of the plane wave accordingly.
\begin{figure}
  \centering
  \includegraphics[width=\figwidth\textwidth]{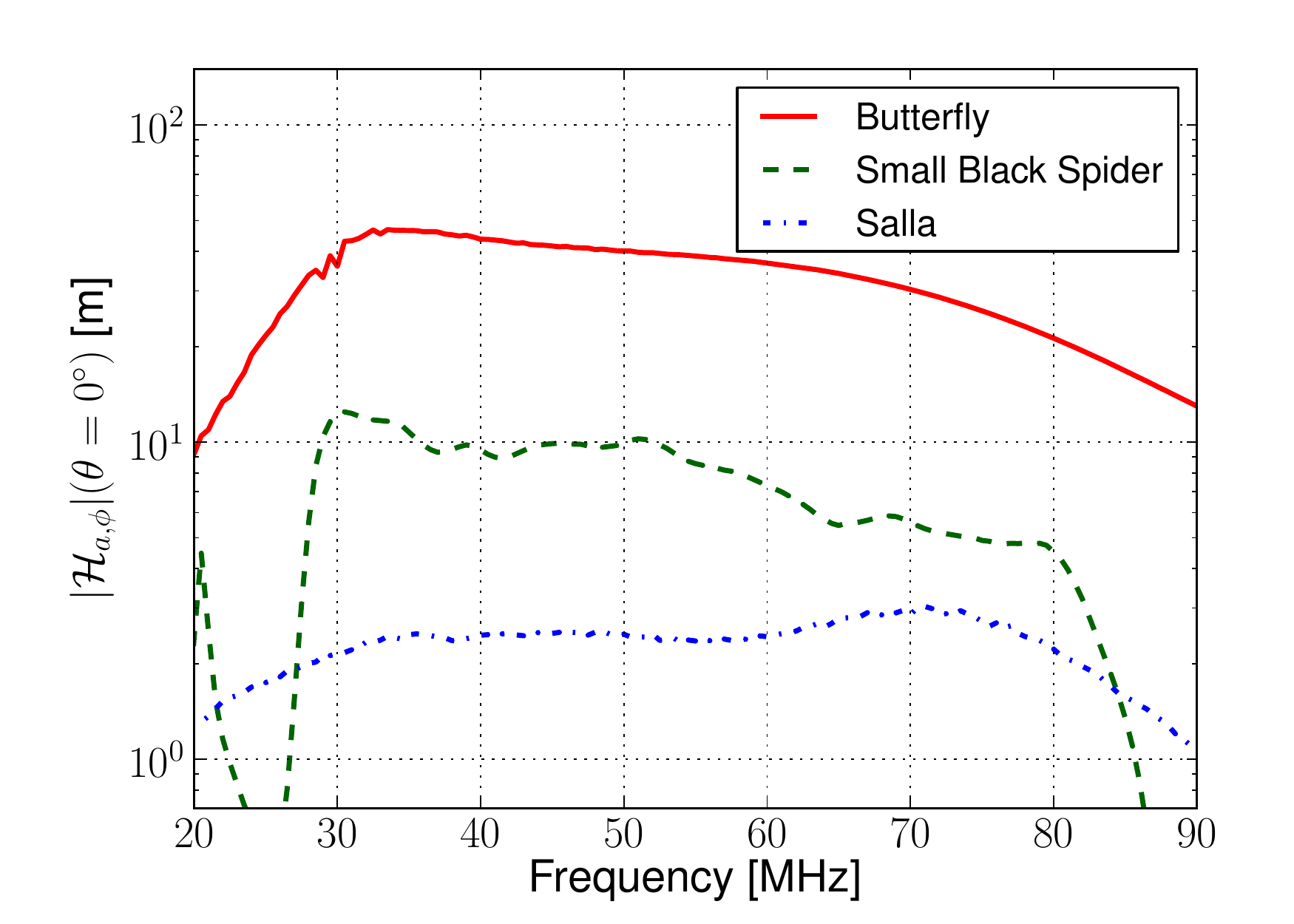}
  \caption{The simulated vector effective length as a function of frequency for the zenith direction for the three tested antennas. Note the logarithmic scale on the y-axis.}
  \label{fig:VELSimulation}
\end{figure}

Alternatively, the tested antenna can be used as transmitter to access the VEL. With the NEC-2 program it is possible to directly compute the far field generated by an antenna at a certain distance $R$. By analogy to Eq. \ref{eqn:EField} the vectorial electric field emitted by an antenna is related to the VEL as:
\begin{equation}
  \label{eqn:reci}
  \vec{\mathcal E}(R) = -i \,\,Z_0 \,\, \frac{1}{2\,\lambda\,R} \,\, \mathcal I_0^t \,\, \vec{\mathcal H} \,\, e^{-i\,\omega\,R/c} \,,
\end{equation}
where $\mathcal I_0^t$ is the current used to operate the antenna. The reciprocity theorem states the equivalence of antenna characteristics obtained from receiving and transmitting measurements. Solving Eq. \ref{eqn:reci} for the VEL yields the desired sensitivity of the antenna to incoming signals. Here it should be noted that the electric field $\vec{\mathcal E'}$ given by the NEC-2 simulation is normalized to a unit distance of $r=1\,\mbox{m}$:
\begin{equation}
  \vec{\mathcal E}'  = \vec{\mathcal E}(R)\,R \, e^{i\,\omega\,R/c} \quad.
\end{equation}

We find that both options yield equivalent results for the VEL $\vec{\mathcal H}$. Using Eqs. \ref{eqn:realizedVELtotal} and \ref{eqn:Vamp} we finally produce $\vec{\mathcal H}_a$ which takes the impact of the amplifier and intermediate transmission lines into account.

\subsection{Characteristics of the Ultra-Wideband Vector Effective Length}
\begin{figure}
  \centering
  \includegraphics[width=\figwidth\textwidth]{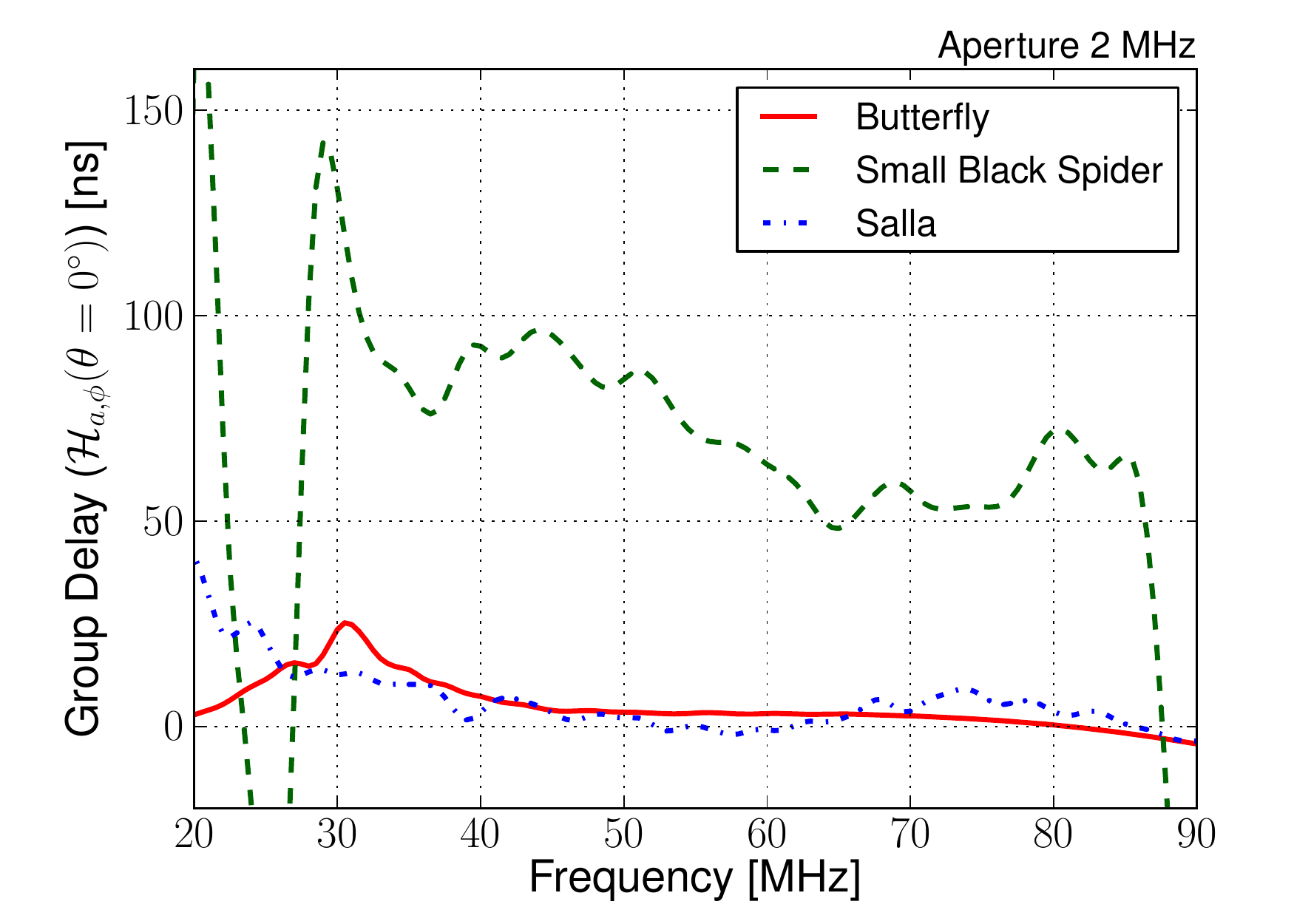}
  \caption{The simulated group delay induced by the antennas for signals coming from the zenith direction $\theta = 0^{\circ}$. A deviation from a non-constant group delay induces dispersion of the measured transient signal.}
  \label{fig:GroupDelaysSimulation}
\end{figure}

In Fig. \ref{fig:VELSimulation} the absolute values of the VELs and in Fig. \ref{fig:GroupDelaysSimulation} the corresponding group delays are displayed as a function of frequency for the three tested antennas.
In the case of the Small Black Spider, the depicted characteristics correspond to the displays in Figs. \ref{fig:GainMeasured} and \ref{fig:GDMeasured} with slight changes due to the adapted simulation setup.

The absolute values of the VEL are rather different for the three antennas. The amplifier of the Butterfly antenna exhibits the strongest gain resulting in large values for the VEL. The amplifiers of the Salla and the Small Black Spider feature similar amplifications. However, the VEL of the Salla is reduced as the antenna is loaded with an ohmic resistor. Note that the average levels of the VELs do not necessarily reflect the signal to noise ratio obtained in measurements. 

Regarding the dependence of the VEL on frequency, both the Butterfly and the Small Black Spider focus on the performance at lower frequencies. The two antennas have been optimized with respect to the frequency content of the radio pulse which is predicted to be governed by longer wavelengths especially when observing air showers at large distances from the shower axis \cite{Ludwig2011,Scholten2008}.

In addition, the Small Black Spider acts as a sharp bandpass to the AERA band. Also, the Butterfly slightly attenuates signals below 30 MHz to avoid short band transmitters. 

The drop-off in sensitivity visible for the Butterfly antenna at the highest frequencies results from the specific observation direction of $\theta = 0^{\circ}$ chosen for the display. Here, the constructive interference of the direct and the wave reflected from the ground is diminished at the height of the Butterfly of 1.5 m above ground for frequencies above 70 MHz.

The Salla antenna has been designed for highest sensitivity at 70 MHz by approaching a situation of conjugate matching between the impedance of the antenna structure and the input impedance of the LNA (cf. Appendix \ref{app:VELandGain}). As will be discussed in Sec. \ref{sec:ObsGal}, the intensity of the radio background noise decreases with frequency. In combination with an increasing antenna sensitivity, the Salla is intended to measure noise spectra that are flat as a function of frequency.

In Fig. \ref{fig:GroupDelaysSimulation} the  group delays of the antennas are pictured. In comparison to the Small Black Spider, the Butterfly antenna and the Salla feature an almost constant group delay within the considered bandwidth from 30 to 80 MHz. 

\subsection{Transient Antenna Characteristics}
Up to now we have discussed the Fourier transform of the VEL $\vec{\mathcal H}_a$. Antenna characteristics inspected at discrete frequencies correspond to the properties of the antenna that will be observed under the reception of mono-frequency signals which implies an infinite signal duration. The transient response is encoded in the development of the antenna characteristics as a function of frequency.

To evaluate the distortion introduced to transient signals by variations of the VEL within the bandwidth we apply the inverse Fourier transform and inspect the VEL in the time domain:
\begin{equation}
  \label{eqn:VELTD}
  \vec{H}_a(t) = \mathcal F^{-1}(\vec{\mathcal H}_a(\omega) ) \quad[\mbox{m Hz}] \quad .
\end{equation}
Examination of the VEL in the time domain combines the full frequency range and takes into account the respective phase relations. Since the bandwidth is limited by the antenna characteristics the treatment of the VEL in the time domain results in a useful estimator.

\begin{figure}
  \centering
  \includegraphics[width=\figwidth\textwidth]{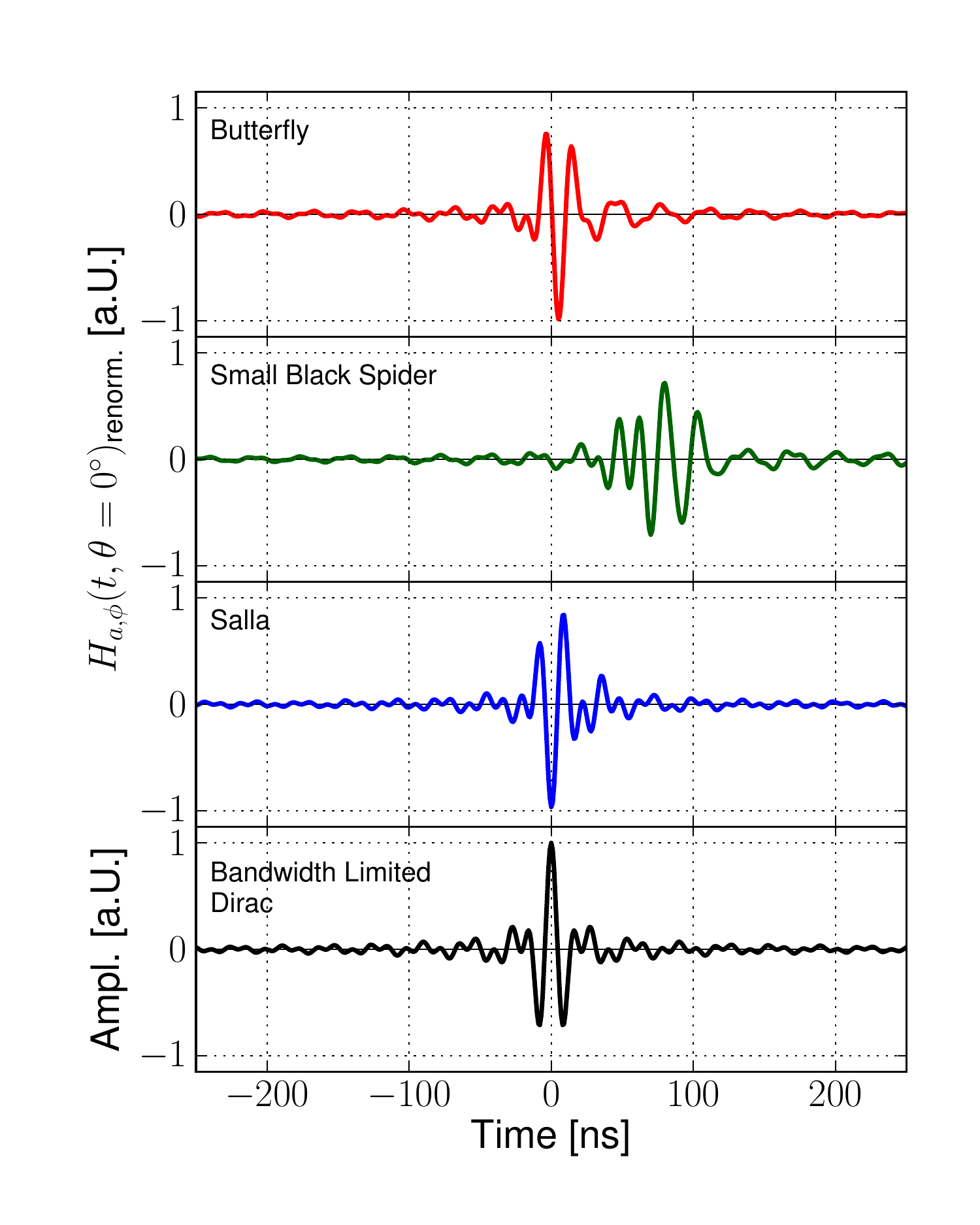}
  \caption{The VEL in the time domain for the zenith direction $\theta = 0^{\circ}$. The VELs have been limited to a common bandwidth of 30 to 80 MHz. The functions are renormalized to their respective maximum peak value which would be realized without dispersion in the antenna. The bandwidth limited Dirac pulse corresponds to an ideal antenna which introduces a bandwidth limitation to incoming signals only.}
  \label{fig:AntennaLengthTD}
\end{figure}

The result of the transformation in Eq. \ref{eqn:VELTD} is displayed in Fig. \ref{fig:AntennaLengthTD} restricted to a common bandwidth of 30 to 80 MHz. In the time domain the antenna characteristics can be interpreted in terms of a transient wave form. The response to an incoming signal is calculated as the convolution of the respective displayed VEL and the incoming wave (cf. Sec. \ref{sec:VectorEffectiveLength}).

For comparison, a bandwidth limited Dirac pulse is also shown. The shape of the Dirac pulse corresponds to a flat transfer function without dispersion. Hence, the Dirac pulse represents the VEL of an antenna that fully reproduces the shape of incoming signals within the bandwidth.

The Salla and the Butterfly antenna 
have an almost unbiased response to incoming signals.
The relatively large average group delay of the Small Black Spider, already indicated in Fig. \ref{fig:GroupDelaysSimulation}, is visible in Fig. \ref{fig:AntennaLengthTD} as a time delay of $H_{a,\phi}(t)$. In comparison to the other antennas the Small Black Spider introduces the largest signal distortion. The pulse is broadened which corresponds to the dispersion induced by the variations of the group delay within the bandwidth.

As normalization to the characteristics displayed in Fig. \ref{fig:AntennaLengthTD} we choose the maximum absolute value of the respective $H_{a,\phi}(t)$ that would be realized if the VEL was not distorted by variations of its group delay. With respect to these maximum values we find that the Salla and the Butterfly antenna keep $> 95$\% of their peak amplitudes whereas the maximum VEL of the Small Black Spider is reduced to $\sim 70$ \% of the undispersed waveform.


In the calibration measurements with the Small Black Spider we have seen that the directional properties of an antenna depend on the frequency under consideration (cf. Fig. \ref{fig:VELvsZen}). Extending the idea in Ref. \cite{Licul2004}, we investigate the maximum peak of the absolute value of the VEL in the time domain as a function of incoming direction. The result is displayed in Fig. \ref{fig:DirectionalPeaks}. The plot is subdivided for the two components of the VEL $H_{a,\theta}$ and $H_{a,\phi}$. In the case of the Small Black Spider, it follows that the function $H_{a,\phi}(t_{\mbox{\footnotesize peak}},\theta,\phi=270^{\circ})$ in the right part of the diagram is a wideband representation of the directional characteristics displayed in Figs. \ref{fig:VELvsZen} and \ref{fig:GDvsZen}.

For the Small Black Spider antenna we find that the side lobe structure is less distinct in the case of the transient characteristic when compared to the side lobes at single frequencies (Fig. \ref{fig:VELvsZen}). The peak response of the Salla antenna is reduced for zenith angles towards $\theta \sim 50 ^{\circ}$ when the $\vec{e}_{\phi}$-component is considered. In this polarization direction the Butterfly antenna features the least complex coverage of the zenith angle range up to $70^{\circ}$. The suppression of zenith angles towards the horizon depends on the heights of the antenna above the ground.

\FIGURE[t]{
  \centering
  \includegraphics[width=0.75\textwidth]{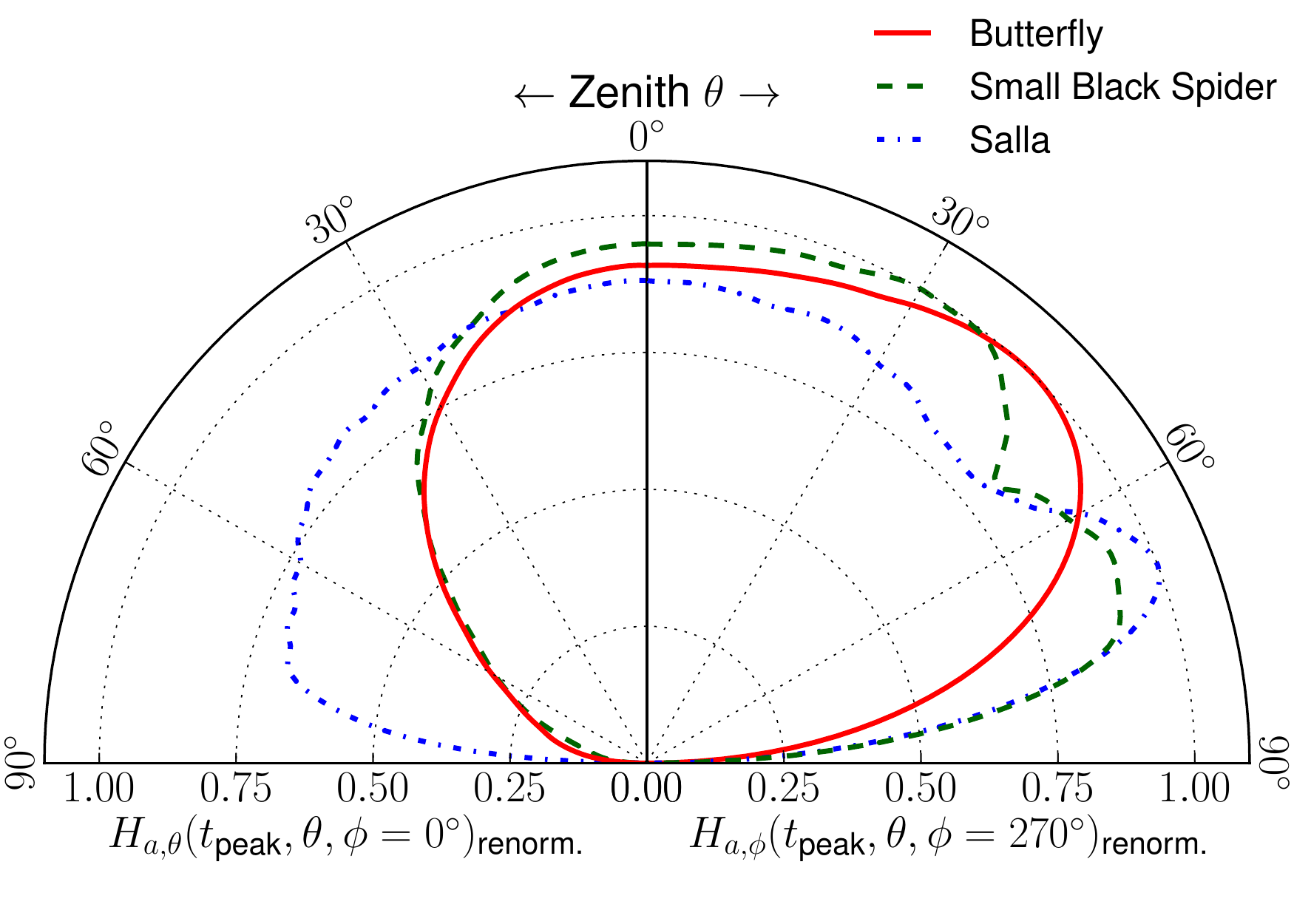}
  \caption{Peak directional diagram: The development of the maximum absolute value of the VEL in the time domain (cf. Fig. \ref{fig:AntennaLengthTD}) is displayed as a function of zenith angle. The two components of the VEL are treated separately. The zenith angle dependence of the $\vec{e}_{\theta}$-component of the VEL is shown in the left part of the diagram. In the right part the $\vec{e}_{\phi}$-component is depicted. Two azimuthal directions have been chosen to maximize the readings of the respective VEL component (cf. Fig. \ref{fig:SBS_gain}). The readings are normalized to the maximum peak value obtained in the zenith angle range. For the zenith direction $\theta = 0^{\circ}$ both components represent the same antenna characteristic which results in a connecting condition for the left and the right part of the diagram. } 
  \label{fig:DirectionalPeaks}
}

In the discussion of Fig. \ref{fig:VectorHeight} it was emphasized that dipole-like antennas become insensitive when the incoming direction of the signal aligns with the dipole axis of the antenna. The development of the $\vec{e}_{\theta}$-component of the peak response is dominated by this geometric effect in the case of the Small Black Spider and the Butterfly antenna, as illustrated in Fig. \ref{fig:DirectionalPeaks}. Here, the circular construction of the Salla antenna leads to a rather constant peak response up to high zenith angles. This results in an enhanced reception of vertically polarized signals from directions close to the horizon.

From the simulation of the antennas we conclude that the Salla and the Butterfly antenna are best suited for the detection of transient signals as they introduce the fewest distortions to the recorded waveform.


\section{Comparison of the Reception of the Galactic Noise Intensity}
\label{sec:ObsGal}
The response voltage to an incoming transient signal needs to be discriminated from a continuous noise floor.
The sensitivity of an antenna can therefore be expressed in terms of signal to noise ratio as obtained from measurements. We have investigated the contribution of the signal to this ratio in the previous sections. To characterize the noise that is recorded with the antennas we have performed dedicated test measurements which are presented in this section.\\

Beyond human-made noise the dominant source of continuous radio signal in the bandwidth from 30 to 80 MHz is the galactic radio background. This diffuse galactic background emission is mainly caused by charged particle gyrating in the magnetic fields of our galaxy \cite{Bridle1967}. 

The spectral irradiance $I_{\nu}(\nu)$ of galactic background depends on the frequency and the observed direction in the sky. The frequency dependence can be described by a power law:
\begin{equation}
  \label{eqn:grb}
  I_{\nu}(\nu) \propto \nu^{-\beta} \quad , \qquad [I] = \mbox{Wm$^{-2}$Hz$^{-1}$} \quad .
\end{equation}
The spectral index $\beta$ ranges from 2.4 to 2.9 and depends on the frequency range and on the observation direction \cite{Lawson1987}.
\begin{figure}
  \centering
  \includegraphics[width=\figwidth\textwidth]{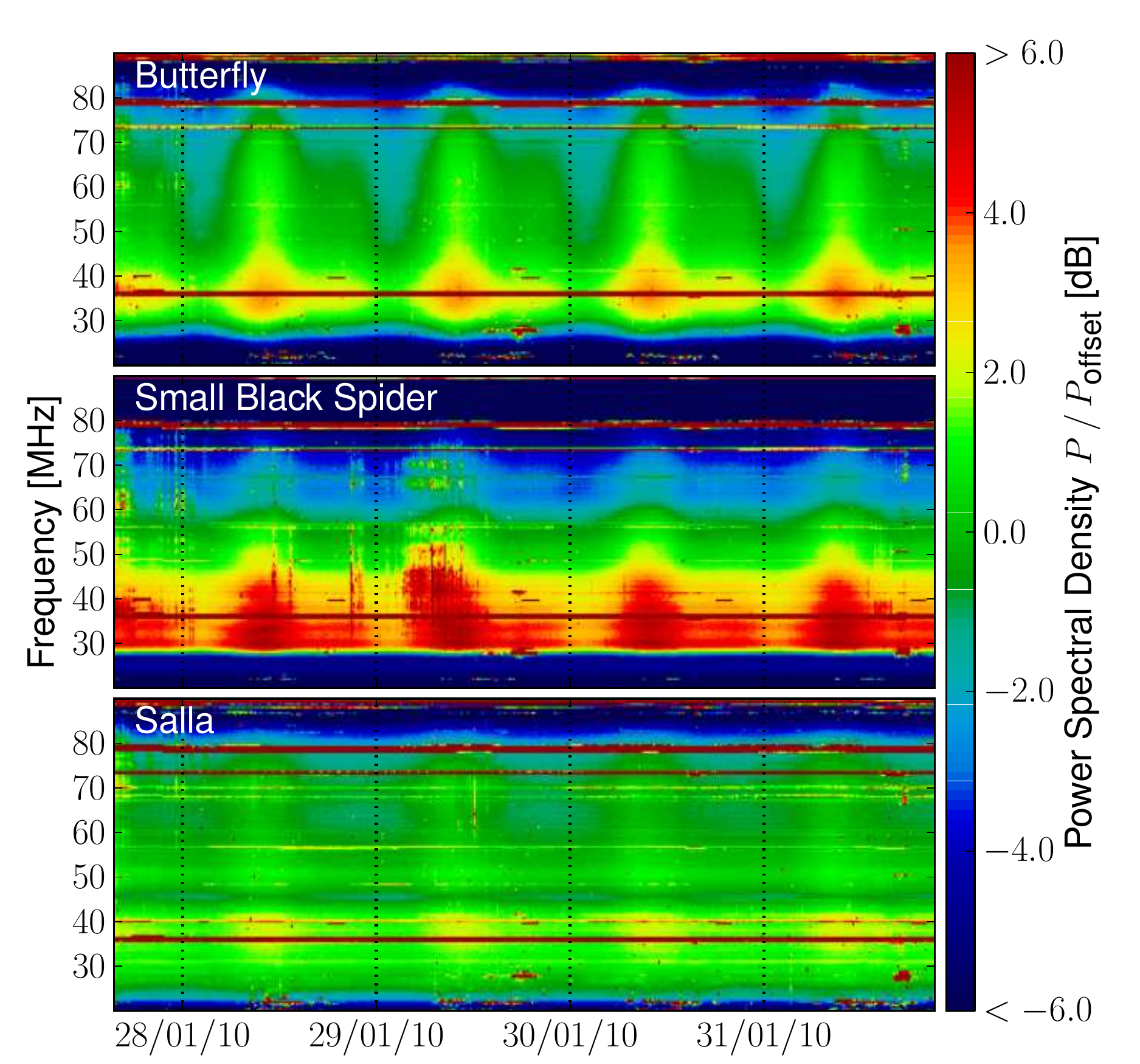}
  \caption{Continuous spectra recorded simultaneously with the three tested antenna at the test bench of Nan\c{c}ay Radio Observatory over a 4 days period. An offset value is chosen for each antenna to allow for a common scale of the power axis. From top to bottom: Butterfly $P_{\mbox{\footnotesize offset}}$ = -50.1 dBm/MHz, Small Black Spider $P_{\mbox{\footnotesize offset}}$ = -62.9 dBm/MHz, Salla $P_{\mbox{\footnotesize offset}}$ = -67.5 dBm/MHz. See text for details.}
  \label{fig:AllDynamicSpectra}
\end{figure}

The dominant contributions to the galactic radio background come from the galactic plane. The tested antennas simultaneously observe a half-sphere of the sky. Consequently, variations of the recorded noise level will occur due to the rise and fall of the galactic plane in the field of view of the antennas. The galactic noise background represents an irreducible background contribution to the observation of air shower signals. However, the variations of the galactic radio signal can be considered as a measure of the antenna sensitivity. These measurements were performed at the Nan\c{c}ay Radio Observatory.

\subsection{Observation of the Galactic Radio Background}
The antenna test bench at the site of the Nan\c{c}ay Radio Observatory is dedicated to the comparison and improvement of antennas used in radio astronomy. Owing to its radio quietness, the Nan\c{c}ay site features similar observing conditions as the site of the Pierre Auger Observatory.

The three antennas under consideration (cf. Sec. \ref{sec:Antennas}) were installed in the test bench for simultaneous observation of the radio background intensity with their read out dipole axes pointing in east-west direction. Besides different coaxial cables, all antennas were read out with the same chain of analog electronics. Frequency spectra from 22 to 82 MHz were recorded from the different antennas in consecutive sweeps with the same spectrum analyzer. We measure one spectrum every $\sim4$ seconds per antenna. Within the measured bandwidth the continuous noise floor of the readout chain is more than 10 dB lower than the average signal power delivered by the Salla antenna, which provides the lowest signal power of the three antennas.

In Fig. \ref{fig:AllDynamicSpectra} the dynamic spectra derived from $10^5$ spectrum sweeps for each antenna are displayed. The readings of the spectrum analyzer are normalized to the resolution bandwidth of the device to yield the recorded power spectral density.

In the dynamic spectra, FM band transmitters appear as horizontal lines at the highest frequencies. At the lowest frequencies, short wave transmitters are visible with varying amplitude due to the changing conditions of the ionosphere. In the relevant frequency range from 30 to 80 MHz only a few transmitters are present, some of them only appearing for short periods of time. Vertical glitches are indicators  for a saturation of the antenna over short periods of time.

For all three tested antennas, a variation over a wide frequency range with a day-like periodicity is visible. This variation results from the changing position of the galactic disc in the field of view of the antennas and follows sidereal time. The sidereal day is $\sim 4$ minutes shorter than the Julian day. 
We have verified the observation of the sidereal period with repeated measurements in the test bench several months apart to rule out man-made interference \cite{Seeger2010}. 

We observe that the variation of the visible galactic radio sky is mapped differently by the antennas onto the recorded power spectral density. Especially in the case of the Salla, the amplitude of the variation appears to be less distinct than for the other antennas. In Sec. \ref{sec:CompGal} we will use the amplitudes to determine the fractions of noise that are added by the antennas internally to the recorded power spectral densities. 

For the site of the Pierre Auger Observatory, long term background observations are described in Ref. \cite{Coppens2009a}.

\subsection{Simulation of Galactic Radio Background Reception}
\FIGURE[t]{
  \centering
  \includegraphics[width=\figwidth\textwidth]{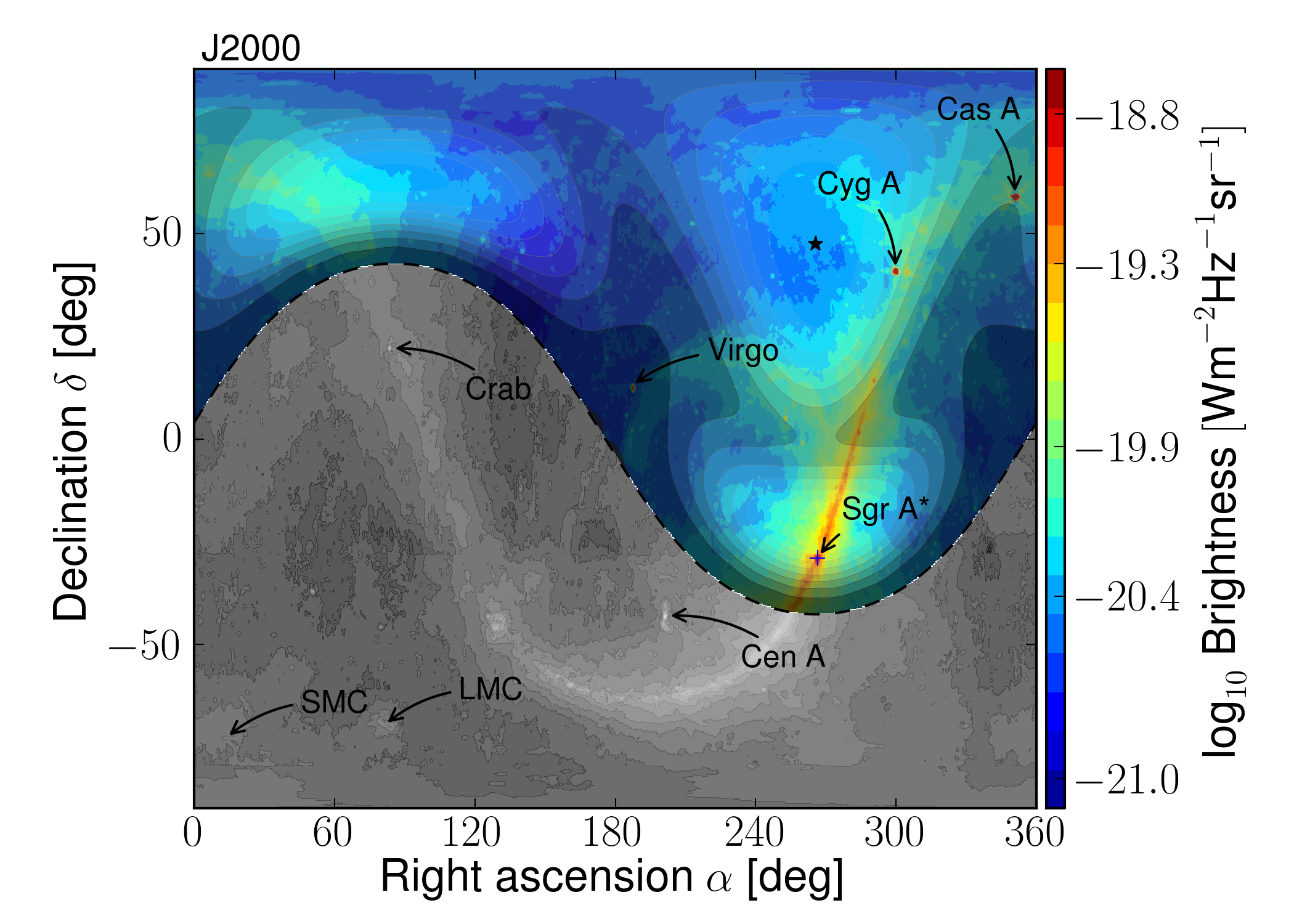}
  \caption{Map of galactic noise intensity generated with LFmap at 55 MHz. Temperatures have been translated to intensities following the Rayleigh-Jeans law. The colored data show noise intensities in the field of view of the Nan\c{c}ay Radio Observatory at a specific time. The horizon is displayed as a dashed line, the star symbol denotes the local zenith direction. The shade over the colored data indicates the relative antenna sensitivity of the Small Black Spider LPDA at the corresponding frequency oriented in east-west direction at Nan\c{c}ay. The measured side lobe in Fig. \ref{fig:VELvsZen} (middle) is here pointing in the direction of the galactic center (Sgr A*). Please refer to a colored version of this plot.}
  \label{fig:P25_SBS_55_}
}
For proper judgment of the measured variations in Fig. \ref{fig:AllDynamicSpectra}, the directional sensitivities of the antennas need to be taken into account. We calculate a prediction of the progression of the power received due to the galactic radio emission using the LFmap program \cite{Polisensky2007}. LFmap combines radio background maps measured at different wavelength and allows for an interpolation to arbitrary frequencies using appropriate power law indices (cf. Eq. \ref{eqn:grb}). We obtain the background brightness $B$ as maps of brightness temperature $T_B$ in equatorial coordinates. The brightness is the spectral irradiance per unit solid angle and is given by the Rayleigh-Jeans law:
\begin{equation}
  B(\nu,\alpha,\delta) = \frac{2 k}{c^2}\,\nu^2 \, T_B \,\quad [B] = \mbox{Wm$^{-2}$sr$^{-1}$Hz$^{-1}$} \, ,
\end{equation}
where $k$ is the Boltzmann constant. In comparison to alternative descriptions of the galactic radio background given by Cane \cite{Cane1979} and the global sky model GSM \cite{Oliveira-Costa2008} we find an agreement of the three models at a level of $\sim 1$ dB by considering the spectral irradiance integrated over the full sky.

An exemplary map of the galactic radio background generated at 55 MHz is displayed in Fig. \ref{fig:P25_SBS_55_}. Besides the brightness, the dashed curve indicates the field of view that contributes to the recorded noise power at the location of the Nan\c{c}ay Radio Observatory at the given time. The relative directionality of the Small Black Spider antenna for the corresponding frequency is represented by the gray shade. 

The noise that is recorded in the measurement is the convolution of the currently visible radio background and the projection of the antenna characteristics onto the sky. The received power spectral density is given by:
\begin{equation}
  \label{eqn:NoisePower}
  P_{\nu}(\nu) = \frac{1}{2}\int\limits_{\Omega} {B(\nu,\alpha,\delta) \, A_{\small e}(\nu,\theta,\phi) \, d\Omega} \quad ,
\end{equation}
where $A_{\small e}$ is the effective aperture of the antenna including mismatch effects and losses (cf. Eq. \ref{eqn:pLoad}).
The factor $1/2$ is introduced explicitly in Eq. \ref{eqn:NoisePower} to take the polarization mismatch between the antenna and an unpolarized radio brightness into account \cite{Kraus1966}. Using the result from Appendix \ref{app:EffectiveAera}, the recorded power spectral density can also be expressed using the VEL:
\begin{equation}
  \label{eqn:NoisePower2}
  P_{\nu}(\nu) = \frac{1}{2} \frac{Z_0}{Z_L}\int\limits_{\Omega} {B(\nu,\alpha,\delta) \, |\vec{\mathcal H}_a(\nu,\theta,\phi)|^2 \, d\Omega}\quad .
\end{equation}

\begin{figure}
  \centering
  \includegraphics[width=\figwidth\textwidth]{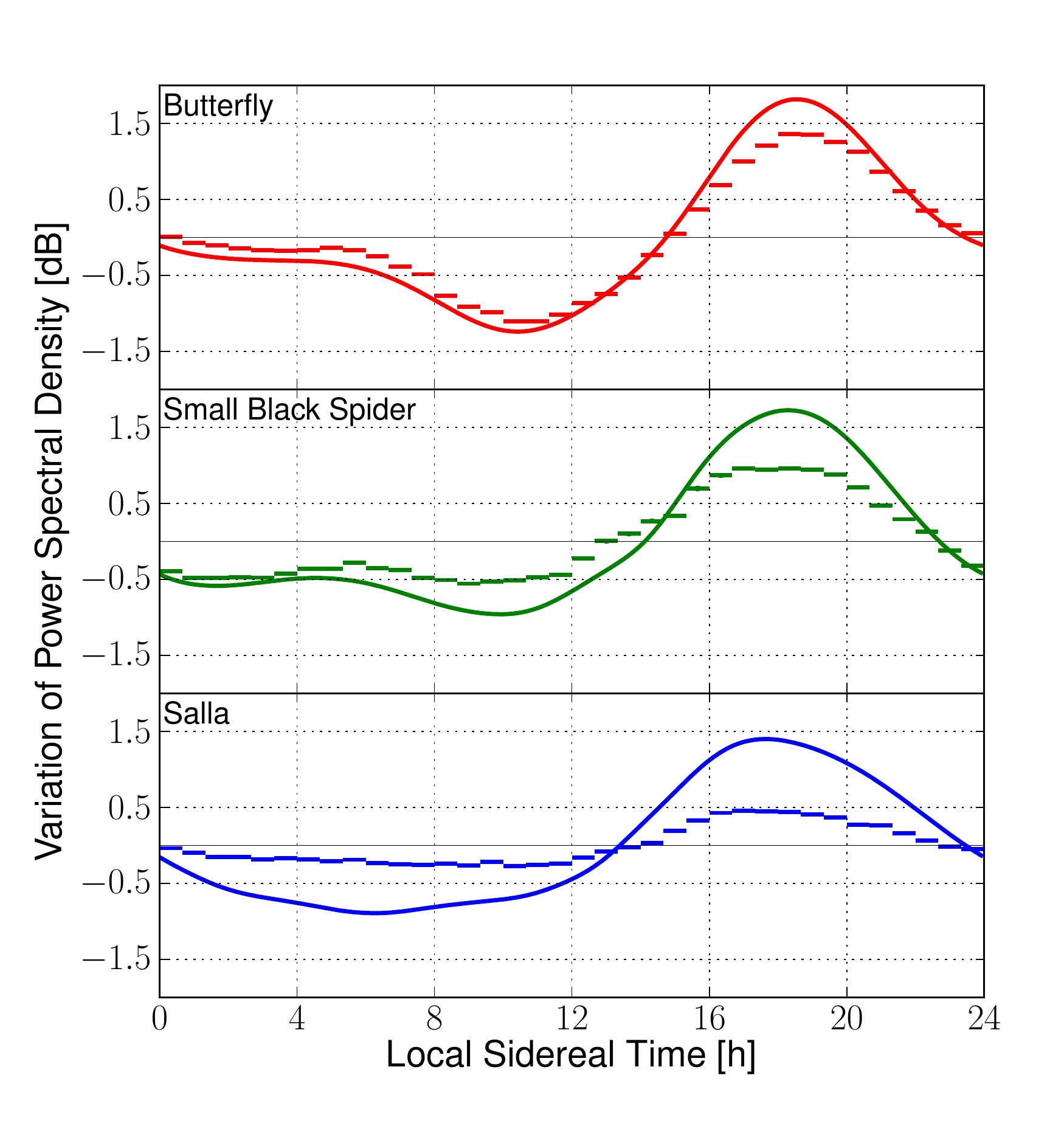}
  \caption{Galactic Variation at the Nan\c{c}ay Radio Observatory at 55 MHz depicted for the Butterfly (top), Small Black Spider (middle) and Salla (bottom). The symbols display the measured data, the solid lines show the corresponding simulations.}
  \label{fig:GalacticVariationRelative55}
\end{figure}
We have compared the measured and simulated average power spectral density as function of frequency based on Eq. \ref{eqn:NoisePower2} using the same simulated VELs as for the discussion of the transient response in Sec. \ref{sec:CompTransResp}. We find that the simulations reproduce the shape of the spectrum. However, they underestimate the noise intensity measured in the setup at the Nan\c{c}ay Radio Observatory by $\sim 4$ dB for all three antennas. For the Small Black Spider we have cross checked the results of the simulation with spectrum observations at the site of AERA. Here, we do not observe differences of the absolute scale of measurements and simulations. In the following section we focus on relative variations of the background noise as a function of sidereal time. Hence, the absolute values of the simulated noise are irrelevant for our further discussion.

\subsection{Comparison of Radio Background Variation}
\label{sec:CompGal}
In Fig. \ref{fig:GalacticVariationRelative55} the variation of the power spectral density as a function of the sidereal time is shown for the measurements and the simulations. 

The measured variation corresponds to a cut in the dynamic spectra of Fig. \ref{fig:AllDynamicSpectra} at 55 MHz where the data have been condensed in the given time binning. For comparison, the measured and the simulated curves are shifted to a common mean value. 

The simulation overestimates the sidereal variation of the noise level without any additional noise sources but the galactic noise. The measured variation of the Butterfly antenna almost follows this ideal curve. If we assume that the simulated variation yields a valid prediction of the noise progression, we conclude that additional external noise sources are negligible for our measurement. At 55 MHz the Butterfly realizes a variation of $\sim2.5$ dB between galactic maximum and minimum where the simulation predicts a maximum variation of $\sim3.1$ dB. For the other two antennas the variation is smaller. 

\begin{figure}
	\centering
	\includegraphics[width=\figwidth\textwidth]{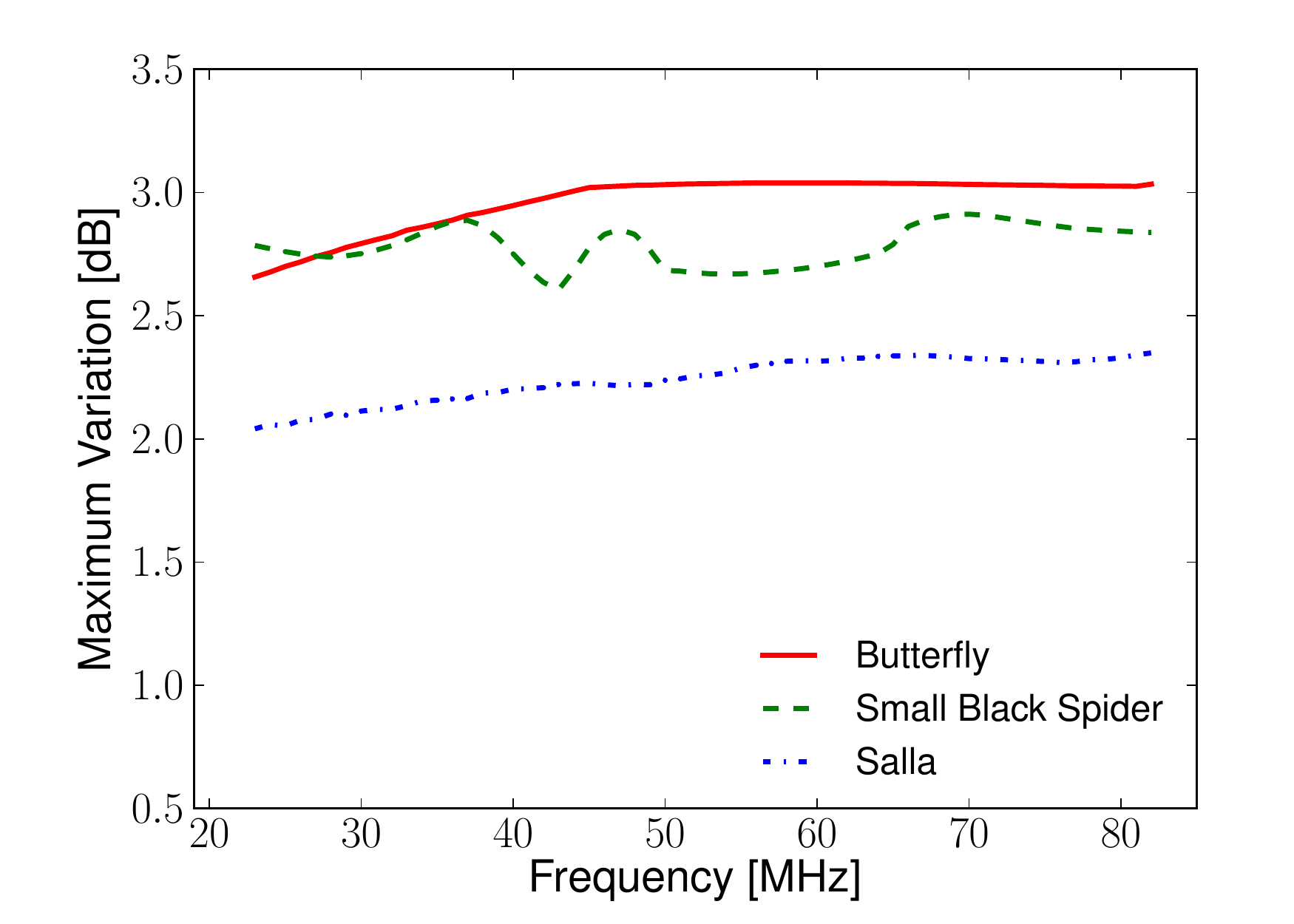}
	\caption{The variation of the galactic noise level obtained from simulations for the three antennas. The Butterfly is predicted to observe the largest difference in received power between galactic maximum and minimum. The variation of the Small Black Spider depends most strongly on the frequency. The galactic noise measured with the Salla is predicted to be least dependent on the time of observation.}
	\label{fig:MaxVar_Sim}
\end{figure}

To evaluate the measured variations with respect to noise that is added in the antenna internally we have to take into account that the predicted maximal variation is different for each antenna. Displayed in Fig. \ref{fig:MaxVar_Sim} is the maximum variation for the full frequency range.

The projection of the directional properties of the Butterfly antenna yields a larger maximal variation than in the case of the other two antennas. The deviations throughout the bandwidth in the case of the Small Black Spider indicate the lobe structure of the antenna evolving with frequency.

The displayed maximum variation $V_{\mbox{\small max}}$ is the ratio of maximal to minimal received galactic noise:
\begin{equation}
  V_{\mbox{\small max}}(\nu) = \frac{P_{\mbox{\small max}}(\nu)}{P_{\mbox{\small min}}(\nu)} \quad .
\end{equation}
For simplicity we assume that the antenna is adding internally an additional noise floor $P_{\mbox{\small int}}$ that is constant with frequency. The realized variation is then:
\begin{equation} 
  \label{eqn:InternalNoise}
  V_{\mbox{\small r}}(\nu) = \frac{P_{\mbox{\small max}}(\nu)+P_{\mbox{\small int}}}{P_{\mbox{\small min}}(\nu)+P_{\mbox{\small int}}} \quad .
\end{equation}
We are interested in the fraction $f$ of internal relative to the galactic noise. In Eq. \ref{eqn:InternalNoise} we count the fraction $f$ relative to the maximum noise that is predicted throughout the whole bandwidth:
\begin{equation} 
  \label{eqn:InternalNoiseFrac}
  V_{\mbox{\small r}}(\nu) = \frac{P_{\mbox{\small max}}(\nu)+f \cdot P_{\mbox{\small max}}(\nu_{\max})}
                                  {P_{\mbox{\small min}}(\nu)+f \cdot P_{\mbox{\small max}}(\nu_{\max})} \quad .
\end{equation}
In Fig. \ref{fig:MaxVarConstantNoiseFloor} the measured variation is displayed as a function of frequency for the three antennas. The simulated variations from Fig. \ref{fig:MaxVar_Sim} have been adjusted with noise fractions $f$ to match the realized variations. 
\FIGURE[t]{
  \centering
  \includegraphics[width=\figwidth\textwidth]{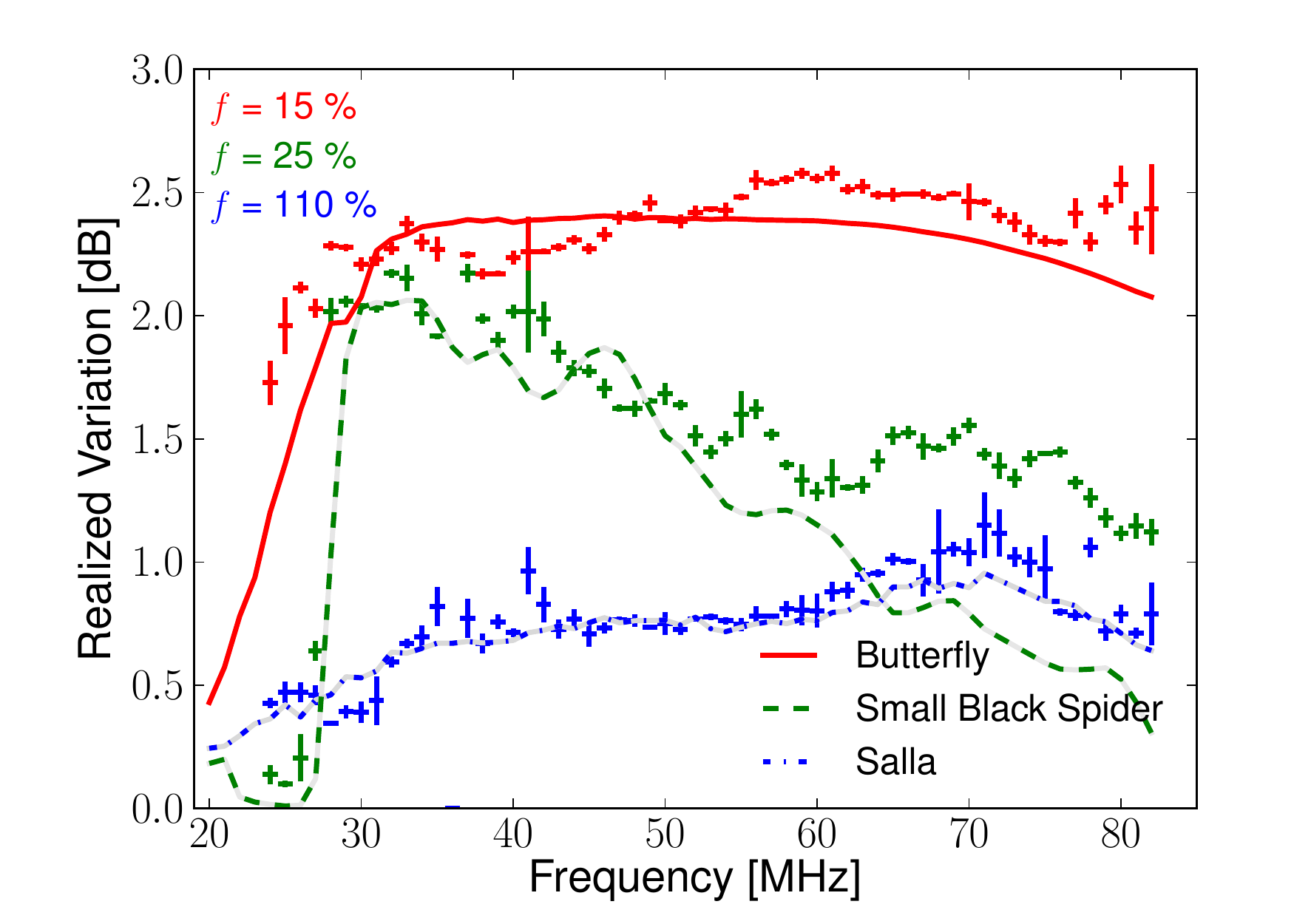}
  \caption{The measured galactic variation for the three antennas as a function of frequency here indicated with the bar symbols. The continuous lines give the expected variation from simulations as in Fig. \ref{fig:MaxVar_Sim} but with a noise power added that is constant in frequency. The percentages denote the strength of the added noise floor relative to the maximum signal expected from the galactic background.}
  \label{fig:MaxVarConstantNoiseFloor}
}

The observed variation of the Salla antenna ranges below 1 dB. Therefore, even at the galactic maximum, more than half of the output power corresponds to internal antenna noise.

The Small Black Spider almost realizes the full variation at the lower frequencies indicating a fraction of internal noise of $\sim25$\%.
However, at the highest frequencies the variation of the galactic background is less distinct. Here, the antenna's sensitivity is not sufficient to raise the galactic signal strongly above the internal noise floor.

The Butterfly antenna provides a power spectral density which is strongly dominated by the galactic noise background over the full frequency range. This constitutes the best observing conditions to air shower signals in terms of continuous background noise.

  \section{Summary and Conclusions}
With AERA --- the Auger Engineering Radio Array --- the Pierre Auger Collaboration addresses both technological and scientific questions of the radio detection of ultra-high energy cosmic rays.

AERA will check the feasibility of the radio-detection technique on large instrumented sites and is thus intended to serve as a blueprint for the next generation of ground-based cosmic ray detectors at very high energies.

The installation of the first stage of AERA (21 stations) is the startup for the construction of a 20 km$^2$ radio detector consisting of 160 autonomous self-triggered detector stations sensitive to frequencies from 30 to 80 MHz. The first stage has been taking data since September 2010.

Along with the construction of the first stage of AERA, we have performed dedicated studies of the antenna acting as sensor to the radio emission. The transient nature of the air shower signal requires a detailed description of the antenna sensor, aiming for a calibrated measurement of the incident signal. We identify the vector effective length as suitable quantity to perform the calculation of the antenna response to transient signals including multiple reflections, interference and polarization effects. 

Having identified the relevant antenna characteristics we reported the calibration of a logarithmic-periodic dipole antenna used at AERA. For the first time, the zenith angle dependency of the sensitivity of a full scale radio detector antenna was accessed including the vectorial phase information. This allowed us to calculate the vector effective length of the antenna on an absolute scale.

For the next setup stage of AERA we have evaluated three candidate antennas each pursuing a different strategy for an optimal reception of cosmic ray signals. A logarithmic-periodic dipole antenna, a loaded dipole loop antenna and an active bowtie antenna were considered.

On the basis of the vector effective length we studied the transient response characteristics of the antennas in the time domain. Our analysis reveals a considerable reduction of the peak amplitude response due to dispersion in the case of the logarithmic-periodic dipole antenna.

At the Nan\c{c}ay Radio Observatory we have performed simultaneous measurements of the galactic radio background with the three antennas. Observing the variation of the measured noise power as a function of sidereal time we have tested the antennas for an optimal continuous noise performance.

We find that the active bowtie antenna is superior in both the noise performance and the transient response. For the next setup stage of AERA we have chosen the bowtie as sensor to the coherent radio emission from cosmic ray induced air showers.


  \acknowledgments
The successful installation, commissioning, and operation of the Pierre Auger Observatory
would not have been possible without the strong commitment and effort
from the technical and administrative staff in Malarg\"ue.

We are very grateful to the following agencies and organizations for financial support: 
Comisi\'on Nacional de Energ\'ia At\'omica, 
Fundaci\'on Antorchas,
Gobierno De La Provincia de Mendoza, 
Municipalidad de Malarg\"ue,
NDM Holdings and Valle Las Le\~nas, in gratitude for their continuing
cooperation over land access, Argentina; 
the Australian Research Council;
Conselho Nacional de Desenvolvimento Cient\'ifico e Tecnol\'ogico (CNPq),
Financiadora de Estudos e Projetos (FINEP),
Funda\c{c}\~ao de Amparo \`a Pesquisa do Estado de Rio de Janeiro (FAPERJ),
Funda\c{c}\~ao de Amparo \`a Pesquisa do Estado de S\~ao Paulo (FAPESP),
Minist\'erio de Ci\^{e}ncia e Tecnologia (MCT), Brazil;
AVCR AV0Z10100502 and AV0Z10100522, GAAV KJB100100904, MSMT-CR LA08016,
LG11044, LC527, 1M06002, MSM0021620859 and RCPTM - CZ.1.05/2.1.00/03.0058, Czech Republic;
Centre de Calcul IN2P3/CNRS, 
Centre National de la Recherche Scientifique (CNRS),
Conseil R\'egional Ile-de-France,
D\'epartement  Physique Nucl\'eaire et Corpusculaire (PNC-IN2P3/CNRS),
D\'epartement Sciences de l'Univers (SDU-INSU/CNRS), France;
Bundesministerium f\"ur Bildung und Forschung (BMBF),
Deutsche Forschungsgemeinschaft (DFG),
Finanzministerium Baden-W\"urttemberg,
Helmholtz-Gemeinschaft Deutscher Forschungszentren (HGF),
Ministerium f\"ur Wissenschaft und Forschung, Nordrhein-Westfalen,
Ministerium f\"ur Wissenschaft, Forschung und Kunst, Baden-W\"urttemberg, Germany; 
Istituto Nazionale di Fisica Nucleare (INFN),
Ministero dell'Istruzione, dell'Universit\`a e della Ricerca (MIUR), Italy;
Consejo Nacional de Ciencia y Tecnolog\'ia (CONACYT), Mexico;
Ministerie van Onderwijs, Cultuur en Wetenschap,
Nederlandse Organisatie voor Wetenschappelijk Onderzoek (NWO),
Stichting voor Fundamenteel Onderzoek der Materie (FOM), Netherlands;
Ministry of Science and Higher Education,
Grant Nos. N N202 200239 and N N202 207238, Poland;
Funda\c{c}\~ao para a Ci\^{e}ncia e a Tecnologia, Portugal;
Ministry for Higher Education, Science, and Technology,
Slovenian Research Agency, Slovenia;
Comunidad de Madrid, 
Consejer\'ia de Educaci\'on de la Comunidad de Castilla La Mancha, 
FEDER funds, 
Ministerio de Ciencia e Innovaci\'on and Consolider-Ingenio 2010 (CPAN),
Xunta de Galicia, Spain;
Science and Technology Facilities Council, United Kingdom;
Department of Energy, Contract Nos. DE-AC02-07CH11359, DE-FR02-04ER41300,
National Science Foundation, Grant No. 0450696,
The Grainger Foundation USA; 
ALFA-EC / HELEN,
European Union 6th Framework Program,
Grant No. MEIF-CT-2005-025057, 
European Union 7th Framework Program, Grant No. PIEF-GA-2008-220240,
and UNESCO.

  \appendix
\gdef\thesection{\Alph{section}}

  \section{Vector Effective Length and Gain}
    \label{app:VELandGain}
    \FIGURE[t]{
      \includegraphics[width=1\textwidth]{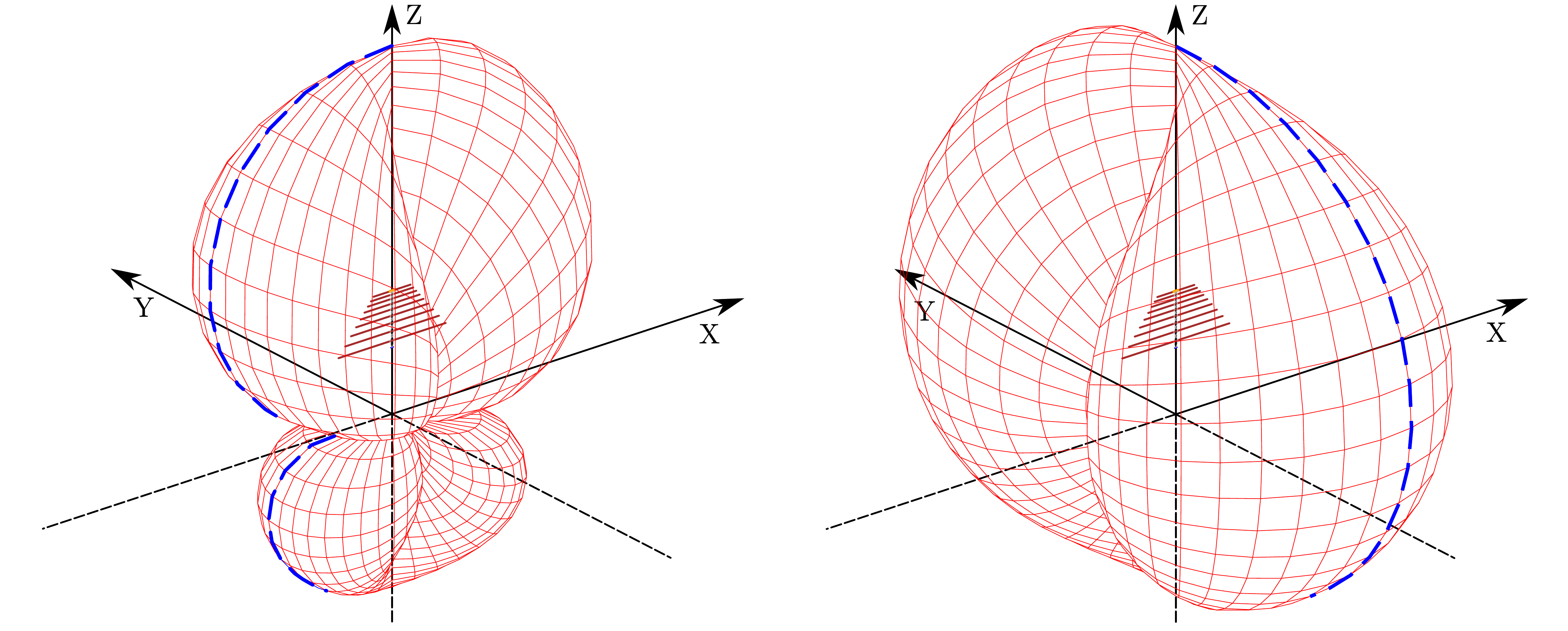}
      \caption{Representation of the vector field of the VEL depicted in Fig. \ref{fig:VectorField} as the antenna gain. For each incoming direction of a signal the gain is given as the distance between the surface and the center of the coordinate system on a logarithmic scale. Left: The components of $\vec{\mathcal H}$ in $\vec{e}_{\theta}$-direction yield the partial gain $G_{\theta}$ which is sometimes called vertical gain. The dashed blue curve presents a cut in the gain sphere which is referred to as E-field plane gain. Right: The components of $\vec{\mathcal H}$ in $\vec{e}_{\phi}$-direction yield the partial gain $G_{\phi}$ also referred to as horizontal gain. Here, the blue dashed curve is called H-field plane gain.}
      \label{fig:SBS_gain}
    }

    According to Eq. \ref{eqn:response}, the VEL allows antenna calculations to be performed on the basis of complex amplitudes. The VEL is related to more commonly used quantities to describe antennas such as the antenna directivity. Directivity is related to the induced signal power. In this section we summarize the interrelation between the power- and the amplitude-based calculations.

    The intensity of a wave averaged over time in a single polarization direction $S_k$ with $k = \theta,\phi$ at a specific frequency is given by:
    \begin{equation}
    \label{eqn:Poynting}
    S_k = \frac{1}{2}\frac{|\mathcal E_k|^2}{Z_0} 
    \end{equation}
    with $Z_0 \approx 120\,\pi\,\Omega$ the vacuum impedance.
    The power that is available to the readout (load, L) of the antenna is then accessible via the maximum effective aperture $A_{\mbox{\small em},k}$ of the antenna:
    \begin{equation}
    \label{eqn:pLoad}
    P_{L,k} = A_{\mbox{\small em},k} \, S_k \quad .
    \end{equation}
    In antenna theory \cite{Balanis2005} the effective aperture is related to the directivity $D$ of the antenna. The directivity for a specific polarization $D_k$ is called partial directivity:
    \begin{equation}
    A_{\mbox{\small em},k} = \frac{\lambda^2}{4\pi}D_k \quad .
    \end{equation}
    Hence the power that is delivered by the antenna structure is:
    \begin{equation}
    P_{L,k} = \frac{\lambda^2}{8\pi}D_k \frac{|\mathcal E_k|^2}{Z_0} \quad .
    \end{equation}
    Using the partial response voltages from Sec. \ref{sec:VectorEffectiveLength} we introduce the VEL into the power calculation:
    \begin{equation}
    \label{eqn:PowerLoad}
    P_{L,k} = \frac{\lambda^2}{8\pi}D_k \frac{|\mathcal V_k|^2/|\mathcal H_k|^2}{Z_0} \quad .
    \end{equation}
    The directivity describes the relative antenna properties to different incoming directions of the signal only. It does not take into account losses inside the antenna structure nor possible mismatches between the antenna and the readout system. The impact of such mismatches is discussed in Sec. \ref{sec:Thevenin}. 

    The power given in Eq. \ref{eqn:PowerLoad} will only be achieved under the condition of conjugate matching. Conjugate matching is discussed in detail in antenna literature as it describes an optimal condition for the reception of signals and simplifies the relevant equations \cite{Balanis2005}. In this case the power available at the readout is related to the open circuit voltage $\mathcal V_{\mbox{\small oc}}$ that is induced by the signal over the footpoint impedance $Z_A$ of the antenna:
    \begin{equation}
    \label{eqn:Conjugate}
    P_{L,k} = \frac{|\mathcal V_{\mbox{\small oc},k}|^2}{8 Re(Z_A)} \quad .
    \end{equation}
    The standards in Ref. \cite{IEEEStd145} define the VEL to map the electric field to the open circuit voltage. Hence we identify $\mathcal V_{\mbox{\small oc},k}$  with the partial voltages $\mathcal V_{\theta}$ and $\mathcal V_{\phi}$ mentioned in Sec. \ref{sec:VectorEffectiveLength}. The combination of Eq. \ref{eqn:PowerLoad} and Eq. \ref{eqn:Conjugate} then yields the relationship between VEL and the directivity:
    \begin{equation}
    \label{eqn:directivity}
    D_k = \frac{Z_0}{Re(Z_A)} \frac{\pi}{\lambda^2} |\mathcal H_k|^2 \quad .
    \end{equation}

    When accessing the VEL in simulations and measurements it is usual to include losses inside the antenna structure e.g. due to ohmic resistance. In terms of power, these losses are accounted for by multiplying the directivity with a dimensionless efficiency factor $\epsilon$. If losses inside the antenna structure are included in the VEL, Eq. \ref{eqn:directivity} relates to the antenna gain $G$:
    \begin{equation}
    \label{eqn:gain}
    G_k = \epsilon\, D_k = \frac{Z_0}{Re(Z_A)} \frac{\pi}{\lambda^2} |\mathcal H_k|^2 \quad .
    \end{equation}
    The treatment of losses inside the antenna structure is important as the loaded loop antenna Salla (cf. Sec. \ref{sec:Antennas}) explicitly introduces an ohmic resistor into its structure to give a specific shape to its gain.

    It should be noted that here the gain is a real number for each of the two polarization directions. Hence the gain is not sufficient to describe subtle effects such as the polarization dependence of the recorded signal as discussed in Sec. \ref{sec:Polarization} or dispersion effects within the antenna structure which invoke a complex phase. To save the concept of a power based calculation, a phase can be introduced in addition to the gain and polarization mismatch factors can be added to the calculation. 

    In Fig. \ref{fig:SBS_gain} the partial gains corresponding to the $\mathcal H_{\theta}$ and the $\mathcal H_{\phi}$ component of the VEL at a single frequency are displayed. Consequently, the two plots present a subset of the information given in Fig. \ref{fig:VectorField}. For a specific direction $(\theta,\phi)$ the magnitude of the gain is displayed as the distance from the center of the coordinate system. The gain spheres given in Fig. \ref{fig:SBS_gain} and subsequent cuts through the patterns are more common representations of antenna characteristics. They are helpful when discussing the directional properties of antennas.

  \section{Reconstruction of the Electric Field in Dual Polarized Measurements}
    \label{app:EFieldReco}
    \begin{figure}
      \centering
      \includegraphics[width=\figwidthS\textwidth]{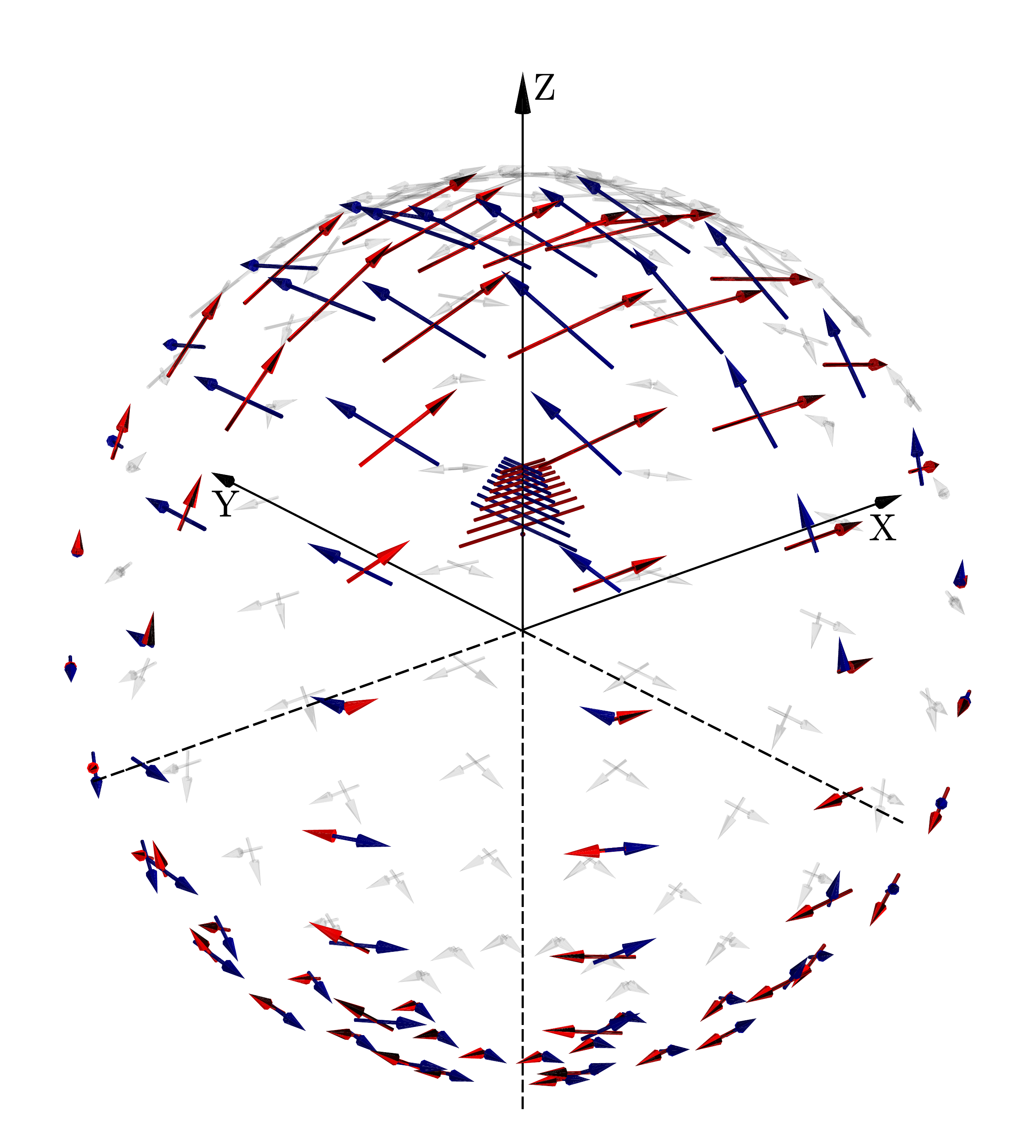}
      \caption{The VEL in the case of a dual polarized antenna setup. Due to two independent measurements of the same incident electric field, the full vectorial signal information can be reconstructed when the incoming direction of the signal is known.}
      \label{fig:DoublePol}
    \end{figure}
    In current radio detection experiments, there are at least two antennas installed at each observing position. In this section we will discuss a major benefit of such dual measurements. That is the reconstruction the 3-dimensional electric field vector of the incoming signal.

    In the case of the antenna models discussed in Sec. \ref{sec:Antennas}, two rotated antennas are assembled in the same hardware structure (cf. Figs. \ref{fig:RDS}, \ref{fig:Salla} and \ref{fig:Butterfly}). Typically, one antenna is rotated by $90^{\circ}$ with respect to the other. In principle, such a setup allows for two independent measurements of the same electric field $\vec{\mathcal E}$ with two different VELs $\vec{\mathcal H}_1$ and $\vec{\mathcal H}_2$. The corresponding vector field of the combined measurement is depicted in Fig. \ref{fig:DoublePol}. It results in two recorded signals that relate as follows:
    \begin{eqnarray}
      \label{eqn:fieldse}
      \mathcal V_1(\omega) &=& \vec{\mathcal H}_1(\omega,\theta,\phi) \cdot \vec{\mathcal E}(\omega) \quad , \notag\\
                          &=& \mathcal H_{1,\theta}(\omega,\theta,\phi) \, \mathcal E_{\theta}(\omega) 
                           \,+\, \mathcal H_{1,\phi}(\omega,\theta,\phi) \, \mathcal E_{\phi}(\omega) 
    \end{eqnarray}
    and
    \begin{eqnarray}
      \label{eqn:fieldse2}
      \mathcal V_2(\omega) &=& \vec{\mathcal H}_2(\omega,\theta,\phi) \cdot \vec{\mathcal E}(\omega) \quad , \notag\\
                          &=& \mathcal H_{2,\theta}(\omega,\theta,\phi) \, \mathcal E_{\theta}(\omega) 
                          \,+\,\mathcal H_{2,\phi}(\omega,\theta,\phi) \, \mathcal E_{\phi}(\omega)\quad .
    \end{eqnarray}
    The system of Eqs. \ref{eqn:fieldse}-\ref{eqn:fieldse2} can be solved for the two components of the electric field vector. To do so, the  antenna characteristics need to be evaluated at the incoming direction $(\theta,\phi)$ of the signal. 

    Here it should be noted that the wave front emitted by an air shower deviates from a planar wave front \cite{Nigl2008}. Hence, the incoming direction of an air shower signal at an antenna sensor is expected to slightly deviate from the direction of the air shower axis. In Ref. \cite{Fliescher2011} we find that this introduces only minor effects in the reconstruction. Changes in the antenna sensitivity are typically small on the scale of angular variations of the incoming direction, which range at a few degree.

    Using, for instance, the direction of the air shower axis the system of Eqs. \ref{eqn:fieldse}-\ref{eqn:fieldse2} can be solved yielding the full electric field vector:
    \begin{eqnarray}
      \label{eqn:fieldsesolved}
      \mathcal E_{\theta}(\omega) 
    &=& \frac{\mathcal V_1(\omega) \, \mathcal H_{2,\phi}(\omega) - \mathcal V_2(\omega) \, \mathcal H_{1,\phi}(\omega)}
              {\mathcal H_{1,\theta}(\omega)\,\mathcal H_{2,\phi}(\omega) - \mathcal H_{1,\phi}(\omega)\,\mathcal H_{2,\theta}(\omega)}  \\
      \mathcal E_{\phi}(\omega)
      &=& \frac{\mathcal V_2(\omega) - \mathcal H_{2,\theta}(\omega) \, \mathcal E_{\theta}(\omega)}
              {\mathcal H_{2,\phi}(\omega)} \quad .
    \end{eqnarray}
    An implementation of this reconstruction scheme is implemented in the `Offline` software framework of the Pierre Auger Observatory \cite{Abreu2011}.

    At this point the response of the dual polarized measurement setup can be expressed in terms of a matrix calculus when the two measured voltage functions in Eq. \ref{eqn:fieldse} are interpreted in terms of a response vector:
    \begin{eqnarray}
      \vec{\mathcal V}(\omega) = \begin{pmatrix} \mathcal H_{1,\theta} & \mathcal H_{1,\phi} \\ \mathcal H_{2,\theta} & \mathcal H_{2,\phi} \end{pmatrix} \cdot \vec{\mathcal E}(\omega) \quad .
    \end{eqnarray}
    For single frequencies the matrix containing the components of the VEL is referred to as Jones antenna matrix or voltage beam matrix in radio polarimetry \cite{J.P.Hamaker1996a}.

  \section{Multiple Reflections in Intermediate Transmission Lines}
  \label{app:IntermediateTransmissionLines}
\begin{figure}
    \centering
    \includegraphics[width=\figwidth\textwidth]{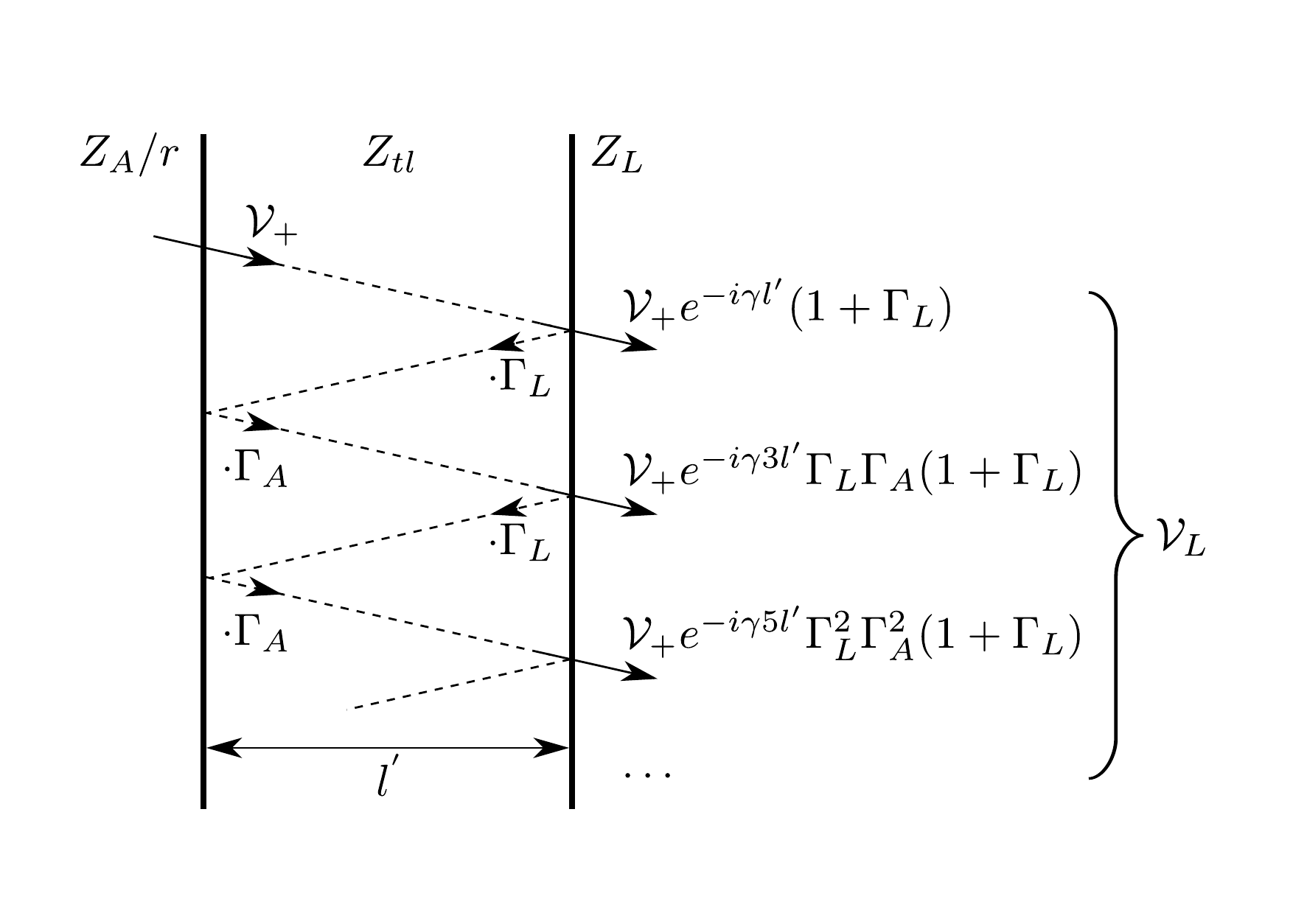}
    \caption{Model for multiple reflection between mismatched source and load impedance with intermediate transmission line.}
    \label{fig:Bounce}
  \end{figure}

  A circuit diagram for a setup with intermediate transmission line is displayed in Fig. \ref{fig:Cir} (right).
  Here the open circuit voltage induces an initial forward traveling voltage $\mathcal V_+$ into the transmission line of impedance $Z_{tl}$. Using the result from Eq. \ref{eqn:Transformer} this voltage is:
  \begin{equation}
    \mathcal V_+ = \frac{\sqrt{r}\, Z_{tl}}{Z_A + r\, Z_{tl}} \, \mathcal V_{\mbox{\small oc}}\quad .
  \end{equation}
  Looking at Fig. \ref{fig:Bounce} we derive a model for the desired voltage amplitude over the readout impedance $\mathcal V_L$. The complex voltage amplitude at the end of the transmission is the initial voltage $\mathcal V_+$ multiplied with a propagation factor:
  \begin{equation}
    \mathcal V_{+} \,(\mbox{at }\,l') = \mathcal V_+ \cdot e^{-(i\frac{\omega}{c} + \alpha) l'} \equiv  \mathcal V_+ \cdot e^{-\gamma l'} \quad ,
  \end{equation}
  where $\gamma$ is the complex propagation constant and $\alpha$ the attenuation loss per unit electrical length $l'$ of the transmission line. According to the Fresnel coefficients, the voltage over the load impedance follows as:
  \begin{equation}
    \label{eqn:VoltageReflection}
    \mathcal V_{0,L} = \mathcal V_+ \cdot e^{-\gamma l'} (1+\Gamma_L) \quad.
  \end{equation}
  Here $\Gamma_L$ is the voltage reflection coefficient:
  \begin{equation}
    \Gamma_L = \frac{Z_L - Z_{tl}}{Z_L + Z_{tl}} \quad .
  \end{equation}
  Following Fig. \ref{fig:Bounce}, part of the amplitude is reflected back towards the antenna. There a further reflection $\Gamma_A$ occurs at the antenna impedance including the transformer operating with the inverted transformation ratio:
  \begin{equation}
    \Gamma_A = \frac{Z_A/r - Z_{tl}}{Z_A/r + Z_{tl}} \quad .
  \end{equation}
  If the scheme in Fig. \ref{fig:Bounce} is continued it yields the total voltage over the load impedance $\mathcal V_L$ as the sum over all amplitudes transferred from the antenna to the load at the end of the transmission line:
  \begin{eqnarray}
  \mathcal V_L &=& \sum_{n=0}^{\infty} \mathcal V_{i,L} \\
                &=& \ldots\,\, = \mathcal V_+ \, (1+\Gamma_L)\, \frac{e^{\gamma l'}}{e^{\gamma 2l'}-\Gamma_A \, \Gamma_L}   \label{eqn:Bounce}  \quad .
  \end{eqnarray}
  In Fig. \ref{fig:AnaVsQucs} the result for the response $V_L(t) = \mathcal F^{-1}(\mathcal V_L(\omega))$ to an exemplary open circuit voltage $V_{\mbox{\small oc}}$ in the time domain is displayed. In this example a setup of $Z_A = Z_L = 200\, \Omega$, $Z_{tl} = 50\, \Omega$ and an electrical line length of $l' = 9\,$m is chosen. These parameters lead to a reappearance of the pulse at the load due to multiple reflections. As a cross check the result for the corresponding setup obtained with the circuit simulator program QUCS \cite{QUCS2009} is displayed showing excellent agreement.

  In summary, combining Eq. \ref{eqn:simpleMatch} and Eq. \ref{eqn:Bounce} we obtain the transfer function that will be realized in a measurement setup with the considered antennas by:
  \begin{equation}
    \rho = \frac{\sqrt{r}\,Z_{tl}}{Z_A + r\,Z_{tl}} \cdot  \frac{(1+\Gamma_L)\,e^{\gamma l'}}{e^{\gamma 2l'}-\Gamma_A\, \Gamma_L}.
  \end{equation}
  \FIGURE[t]{
    \centering
    \includegraphics[width=\figwidth\textwidth]{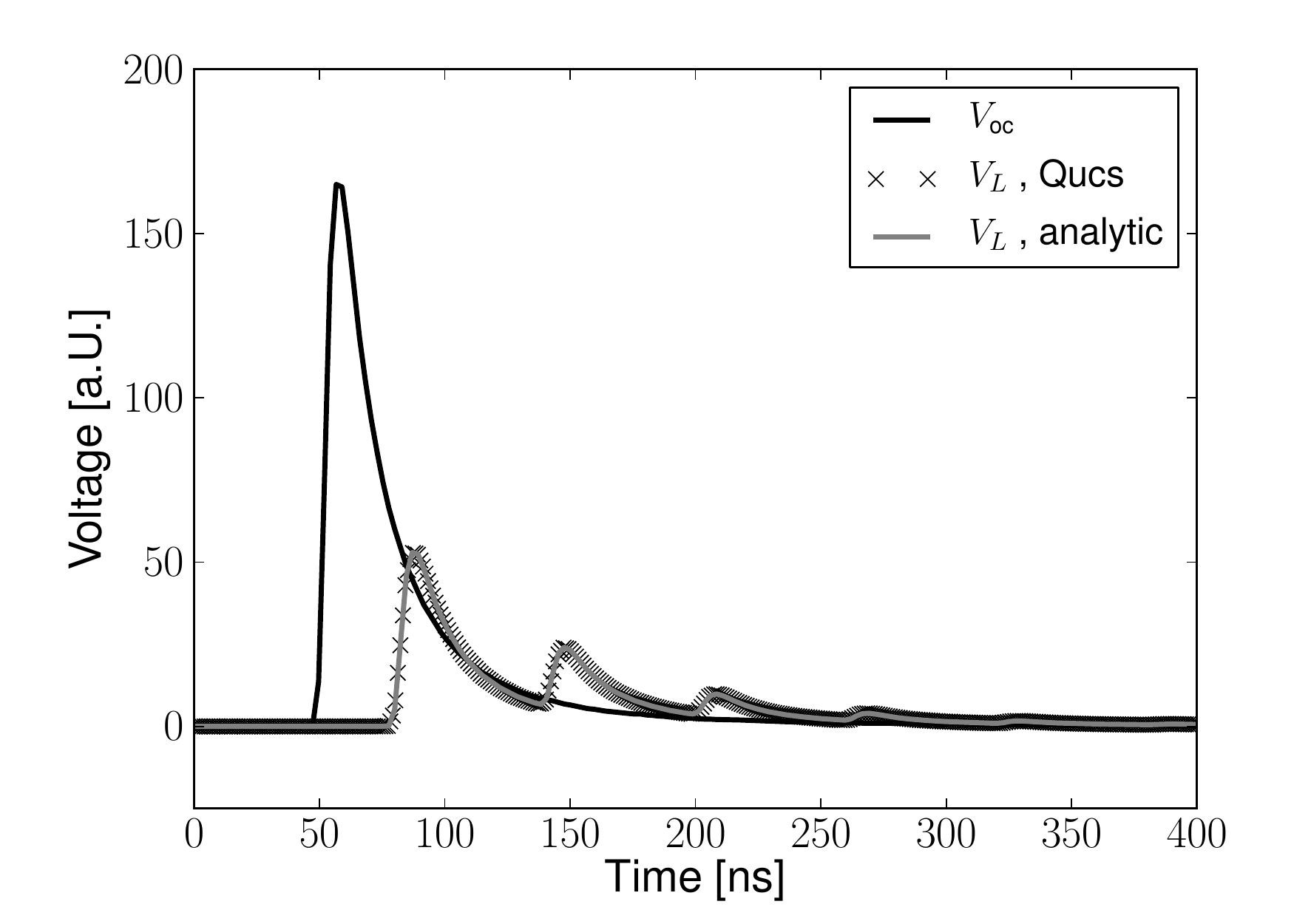}
    \caption{Analytic calculation of the voltage $V_L$ over the load impedance as a response to an initial voltage pulse $V_{\mbox{\small oc}}$: for the circuit diagram depicted in Fig. \ref{fig:Cir} (right). The calculation is performed with the ansatz presented in Fig. \ref{fig:Bounce}. The result is compared to the response simulated with the circuit simulator QUCS \cite{QUCS2009}.
    }
    \label{fig:AnaVsQucs}
  }

  \section{Renormalization of S-Parameter \mbox{S}21}
  \label{app:Sparameter}
  When characterizing an amplifier with scattering parameters, the S-Parameter \mbox{S}21 yields the amplified voltage amplitude normalized to the incoming voltage amplitude $\mathcal V_{g}$ delivered by the signal generator:
  \begin{equation}
    \mbox{S}21 = \frac{\mathcal V_{a}}{\mathcal V_{g}} \quad .
  \end{equation}
  The realized VEL discussed in section \ref{sec:RVEH} relates the incoming electric field to the voltage over the input impedance of the amplifier $\mathcal V_L$. From the S-parameter measurement the voltage at the amplifier input is calculated analog to Eq. \ref{eqn:VoltageReflection} using the voltage reflection coefficient $\Gamma_L$:
  \begin{equation}
    \mathcal V_L = \mathcal V_g\,(1+\Gamma_L) =  \mathcal V_g\,(1+\mbox{S}11) \,
  \end{equation}
  where the reflection coefficient corresponds to the scattering parameter $\mbox{S}11$. Hence, the amplification of the response of the antenna to an electric field calculated with the VEL is given as:
  \begin{equation}
    \frac{\mathcal V_a}{\mathcal V_L} = \frac{\mathcal V_a}{\mathcal V_g (1+\mbox{S}11)}  = \frac{\mbox{S}21}{1+\mbox{S}11} \equiv \mbox{S}21' \quad .
  \end{equation}

  \section{Vector Effective Length and Realized Gain}
  \label{app:velandrelgain}
  The active component of the power that is consumed by an antenna used as transmitter is given by:
  \begin{equation}
    \label{eqn:transpower}
    P^t = \frac{|\mathcal I_0^t|^2}{2} \cdot Re(Z_a^t) \quad .
  \end{equation}
  $\mathcal I_0^t$ is the current within the antenna structure analog to the receiving case depicted in Fig. \ref{fig:Cir}. $Z_a^t$ is the impedance of the transmitter. The power that is delivered through the coaxial cable to the transmitter is: 
  \begin{equation}
    P_g = \frac{1}{Z_{tl}}\frac{|\mathcal V_g|^2}{2} \quad ,
  \end{equation}
  where the voltage amplitude $\mathcal V_g$ is the denominator of the S-parameter \mbox{S}21. Due to the impedance mismatch between the transmitter antenna and the coaxial cable, only a fraction of the power $P_g$ is available to be consumed by the antenna:
  \begin{equation}
    \label{eqn:app_i0}
    |\mathcal I_0^t|^2 \cdot Re(Z_a^t) = \frac{|\mathcal V_g|^2}{Z_{tl}} (1-|\Gamma^t|^2) \quad,
  \end{equation}
  with $\Gamma^t$ the voltage reflection coefficient of the transmitting antenna to the coaxial cable. We resort Eq. \ref{eqn:app_i0} and multiply with the squared VEL of the transmitter:
  \begin{equation}
    \label{eqn:app_vel}
    \frac{|\mathcal I_0^t|^2}{|\mathcal V_g|^2} \, |\mathcal H_{\phi}^t|^2 = \frac{1}{Re(Z_a^t)Z_{tl}} (1-|\Gamma^t|^2) \, |\mathcal H_{\phi}^t|^2 \quad . 
  \end{equation}
  In Eq. \ref{eqn:gain} we have derived the relationship between VEL and gain. Introducing this interrelation on the right hand side of Eq. \ref{eqn:app_vel} yields:
  \begin{equation}
    \frac{|\mathcal I_0^t|^2}{|\mathcal V_g|^2} \, |\mathcal H_{\phi}^t|^2  = 
    \frac{\lambda^2}{\pi Z_0 Z_{tl}}\,\underbrace{(1-|\Gamma^t|^2) G^t}_{G_{\mbox{\footnotesize cal}}^t} \quad ,
  \end{equation}
  which is the result claimed in Eq. \ref{eqn:app1}. The calibration of the transmitter antenna $G_{\mbox{\footnotesize cal}}^t$ is given by the manufacturer including the reflection at the input of the antenna. 

  \section{Effective Aperture and Vector Effective Length}
  \label{app:EffectiveAera}
  In Eq. \ref{eqn:pLoad} the power available to the load due to an incident wave is given in terms of the maximum effective aperture of the antenna. The maximum effective aperture is obtained under conjugate matching and without losses in the antenna. If losses and mismatching effects are included we refer to the effective aperture $A_{\mbox{\small e}}$ as:
  \begin{equation}
    \label{eqn:pLoad2}
    P_{L} = A_{\mbox{\small e}} \, S \quad .
  \end{equation}
  The available power is also accessible via the realized vector effective length:
  \begin{equation}
    \label{eqn:PLPower}
    P_L = \frac{1}{Z_L}\frac{|\mathcal V_L|^2}{2} = \frac{1}{Z_L}\frac{|\vec{\mathcal H}_r \, \vec{\mathcal E} |^2}{2}
  \end{equation}
  The vector product in Eq. \ref{eqn:PLPower} is the projection of the electric field vector onto the vector of the VEL. To disentangle the vectorial from the power calculation a polarization factor $p$ is introduced \cite{Anderson2005}:
  \begin{equation}
    \label{eqn:polFac}
    p = \frac{|\vec{\mathcal H}_r \, \vec{\mathcal E}|^2}{|\vec{\mathcal H}_r|^2 |\vec{\mathcal E}|^2} \quad ,
  \end{equation}
  such that:
  \begin{equation}
    \label{eqn:PLPower2}
    P_L = p \cdot \frac{1}{Z_L}\frac{|\vec{\mathcal H}_r|^2 \,|\vec{\mathcal E}|^2}{2} \quad .
  \end{equation}
  Using Eq. \ref{eqn:Poynting} we introduce the intensity of the wave which is also used in Eq. \ref{eqn:pLoad2}:
  \begin{equation}
    \label{eqn:PLPower3}
    P_L = p \cdot \frac{Z_0}{Z_L}|\vec{\mathcal H}_r|^2 \,S \quad .
  \end{equation}
  Hence we derive the relation between realized VEL and the effective aperture:
  \begin{equation}
    A_{\mbox{\small e}} = p\frac{Z_0}{Z_L} |\vec{\mathcal H}_r|^2 \, .
  \end{equation}
  For an unpolarized signal the electric field vector and the vector of the effective length will align randomly in time. In time averaged measurements the expectation value of the polarization factor is hence:
  \begin{equation}
    <p> = 1/2 \quad .
  \end{equation}

  \section{List of Acronyms}
  \begin{acronym}[EASIERxxxxx]
  \renewcommand{\bflabel}[1]{\normalfont{\normalsize{#1}}\hfill}
  \acro{AERA}{Auger Engineering Radio Array}
  \acro{AM}{Amplitude Modulation}
  \acro{ASIC}{Application-Specific Integrated Circuit}
  \acro{CODALEMA}{COsmic ray Detection Array with Logarithmic ElectroMagnetic Antennas}
  \acro{FM}{Frequency Modulation}
  \acro{FPGA}{Field-Programmable Gate Array}
  \acro{GPS}{Global Positioning System}
  \acro{LMC}{Large Magellanic Cloud}
  \acro{LNA}{Low Noise Amplifier}
  \acro{LOFAR}{LOw Frequency ARray}
  \acro{LOPES}{LO\scriptsize{FAR} \normalsize{PrototypE Station}}
  \acro{LOPES-STAR}{LOPES Self Triggered Array of Radio detectors}
  \acro{LPDA}{Logarithmic-Periodic Dipole Antenna}
  \acro{NEC}{Numerical Electromagnetics Code}
  \acro{SALLA}{Short Aperiodic Loaded Loop Antenna}
  \acro{SMC}{Small Magellanic Cloud}
  \acro{VEL}{Vector Effective Length}
  \acro{QUCS}{Quite Universal Circuit Simulator}
  \end{acronym}  
  \bibliographystyle{JHEP.bst}
  \bibliography{library_antennapaper}

%
%
%
%
%

\end{document}